\documentclass[prx,twocolumn,floatix,amssymb,amsmath, superscriptaddress,longbibliography]{revtex4-2}
\usepackage{graphicx}
\usepackage{epsfig}
\usepackage{amsmath,amssymb,amsthm}
\usepackage[utf8]{inputenc}
\usepackage[dvipsnames]{xcolor}
\usepackage[colorlinks, urlcolor=blue, linkcolor=red]{hyperref}
\usepackage{soul,xcolor,braket}
\usepackage{comment}

\setstcolor{red}

\newcommand{\beqa}{\begin{eqnarray}}
\newcommand{\eeqa}{\end{eqnarray}}

\newcommand{\crea}[1]{{\hat #1}^\dagger}
\newcommand{\ann}[1]{\hat #1}

\renewcommand{\boxed}[2]{\textcolor{#1}{%
\tikz[baseline={([yshift=-1ex]current bounding box.center)}] \node [rectangle, minimum width=1ex,rounded corners,draw] {\normalcolor\m@th$\displaystyle#2$};}}

\newcounter{appsection}
\newcounter{appsubsection}[appsection]

\usepackage{physics}
\usepackage{bookmark}
\usepackage[normalem]{ulem}

\newcommand\redsout{\bgroup\markoverwith{\textcolor{red}{\rule[0.5ex]{2pt}{2pt}}}\ULon}
\newcommand\bluesout{\bgroup\markoverwith{\textcolor{blue}{\rule[0.5ex]{2pt}{2pt}}}\ULon}
\newcommand\greensout{\bgroup\markoverwith{\textcolor{green}{\rule[0.5ex]{2pt}{2pt}}}\ULon}
\newcommand{\overbar}[1]{\mkern 1.5mu\overline{\mkern-1.5mu#1\mkern-1.5mu}\mkern 1.5mu}

\newtheorem{prop}{Proposition}

\begin{document}

\title{Waveguide QED with Quadratic Light-Matter Interactions}
\author{Uesli Alushi}
\affiliation{Department of Information and Communications Engineering, Aalto University, Espoo 02150, Finland}
\author{Tom\'as Ramos}
\affiliation{Institute of Fundamental Physics IFF-CSIC, Calle Serrano 113b, 28006 Madrid, Spain}
\author{Juan Jos\'e Garc\'ia-Ripoll}
\affiliation{Institute of Fundamental Physics IFF-CSIC, Calle Serrano 113b, 28006 Madrid, Spain}
\author{Roberto Di Candia}
\affiliation{Department of Information and Communications Engineering, Aalto University, Espoo 02150, Finland}
\affiliation{Dipartimento di Fisica, Universit\`a degli Studi di Pavia, Via Agostino Bassi 6, I-27100, Pavia, Italy}
\author{Simone Felicetti}
\affiliation{Institute for Complex Systems, National Research Council (ISC-CNR) and Physics Department, Sapienza University, P.le A. Moro 2, 00185 Rome, Italy}

\begin{abstract}
Quadratic light-matter interactions are nonlinear couplings such that quantum emitters interact with photonic or phononic modes exclusively via the exchange of excitation pairs. Implementable with atomic and solid-state systems, these couplings lead to a plethora of phenomena that have been characterized in the context of cavity QED, where quantum emitters interact with localized bosonic modes. Here, we explore quadratic interactions in a waveguide QED setting, where quantum emitters interact with propagating fields confined in a one-dimensional environment. We develop a general scattering theory under the Markov approximation and discuss paradigmatic examples for spontaneous emission and scattering of biphoton states. Our analytical and semi-analytical results unveil fundamental differences with respect to conventional waveguide QED systems, such as the spontaneous emission of frequency-entangled photon pairs or the full transparency of the emitter to single-photon inputs. This unlocks new opportunities in quantum information processing with propagating photons. As a striking example, we show that a single quadratically-coupled emitter can implement a two-photon logic gate with unit fidelity, circumventing a no-go theorem derived for conventional waveguide-QED interactions.
\end{abstract}

\maketitle

\section{Introduction}
The study of light-matter interactions is one of the main research pillars of quantum science. The implementation of systems where quantum emitters interact strongly with confined modes of the electromagnetic fields has allowed us to achieve an unprecedented level of control over quantum degrees of freedom~\cite{HarocheBook,QuantumNoise}. In the framework of cavity quantum electrodynamics (QED), the confinement of the electromagnetic field makes it possible to observe a coherent exchange of excitations between \emph{localized} photonic modes and single quantum emitters~\cite{chang_quantum_2014}. 
Similarly, the coupling of quantum emitters to \emph{propagating} fields can be strongly enhanced using waveguide structures, which confine photons to a one-dimensional environment~\cite{chang_colloquium_2018,sheremet_waveguide_2022}. This setup, known as waveguide QED, has been realized in a variety of platforms such as atoms~\cite{Rauschenbeutel,goban_superradiance_2015,solano_super-radiance_2017,corzo_waveguide-coupled_2019} or quantum dots~\cite{Arcari2014,grim_scattering_2022} coupled to photonic waveguides, as well as superconducting qubits~\cite{astafiev_resonance_2010,forn-diaz_ultrastrong_2017,liu_quantum_2017,eder_quantum_2018,mirhosseini_cavity_2019,kannan_waveguide_2020,brehm_waveguide_2021,scigliuzzo_controlling_2022,zanner_coherent_2022} coupled to microwave transmission lines. Waveguide QED structures have a great potential to implement building blocks of quantum networks~\cite{lodahl_quantum-dot_2017,kimble_quantum_2008}, since propagating photons are ideal to transport flying qubits over long distances while  emitters can provide the strong quantum non-linearity necessary for quantum information processing. Therefore, there has been intense theoretical~\cite{fan_input-output_2010,del_valle_theory_2012,GonzalezTudela15,shi_multiphoton-scattering_2015,ramos_multiphoton_2017,ramos_correlated_2018,Mahmoodian20,trivedi_generation_2020} and experimental~\cite{Stiesdal18,prasad_correlating_2020,le_jeannic_experimental_2021,kannan_generating_2020,Hinney21,le_jeannic_dynamical_2022} research in the control and characterization of few-photon correlations generated by single quantum emitters, as well as in the implementation of photonic devices working at the few-photon level~\cite{hoi_demonstration_2011,bennett_semiconductor_2016,scheucher_quantum_2016,nogo_theorem,uppu_quantum-dot-based_2021,reuer_realization_2022,Yang22}.

In the vast majority of platforms, quantum emitters couple linearly, via dipolar interactions, to photonic or phononic modes. Such interactions support only the exchange of individual quanta---e.g.~an atom decays emitting a single photon---. Transitions involving multiple quanta appear as higher-order processes and are therefore strongly suppressed. Only recently, it has been shown how to implement nonlinear couplings where a quantum emitter interacts with localized bosonic modes via the direct exchange of two excitation quanta (e.g. an atom decays emitting a photon pair). Often dubbed \emph{two-photon} interactions, these quadratic light-matter couplings have been proposed in cavity QED settings such as superconducting qubits non-linearly coupled to quantum microwave resonators~\cite{felicetti_two-photon_2018,PhysRevA.98.053859,bertet2005dephasing} or nanomechanical oscillators~\cite{Zhou2006}. Alternatively, quadratic couplings can also be effectively induced via parametric pumping and other simulation schemes intrinsic to superconducting circuits~\cite{Steele_sidebands}, hybrid spin-nanomechanical oscillators~\cite{Wang2016,Munoz2018}, trapped ions~\cite{felicetti_spectral_2015,PhysRevA.97.023624,PhysRevA.99.032303,Cong_selective}, and ultracold atoms~\cite{Schneeweiss_2018, Dareau_2018}. Notice that non-dipolar~\cite{PhysRevLett.121.060503} couplings have already been observed using superconducting artificial atoms.

The rich quantum phenomenology arising in quadratic light-matter interactions   motivates a fast-growing interest. For example, the two-photon quantum Rabi model is characterized by counter-intuitive spectral  features such as the spectral collapse~\cite{PhysRevA.85.043805,duan2016two,PhysRevA.99.013815,Xie21}, which can have direct observable consequences~\cite{felicetti_spectral_2015,PhysRevA.98.053859}.
Strong quadratic emitter-field couplings can induce high-order quantum optical nonlinear processes~\cite{felicetti_two-photon_2018,zou2019multiphoton,Wang2021}. Quantum phase transitions~\cite{garbe_superradiant_2017,PhysRevA.97.053821,PhysRevA.100.033608,garbe2020dissipation,Cui2020,Ying21,li2022nonlinear} and quantum collective-emission phenomena~\cite{Delmonte_2ph_battery,Piccione2022} have also been analyzed. This phenomenology can be exploited in quantum-information applications such as non-classical state generation~\cite{PhysRevLett.122.123604,casanova2018connecting,PhysRevA.99.023854}, quantum sensing~\cite{Ying_Metrology}, cat-qubit stabilization~\cite{TPE_Cat} and qubit-noise spectroscopy~\cite{mutter_transmission-based_2023}. However, so far quadratic light-matter couplings have only been studied in the case of \emph{localized} bosonic modes.

\begin{figure}[ht!]
\center
\includegraphics[width=.9\linewidth]{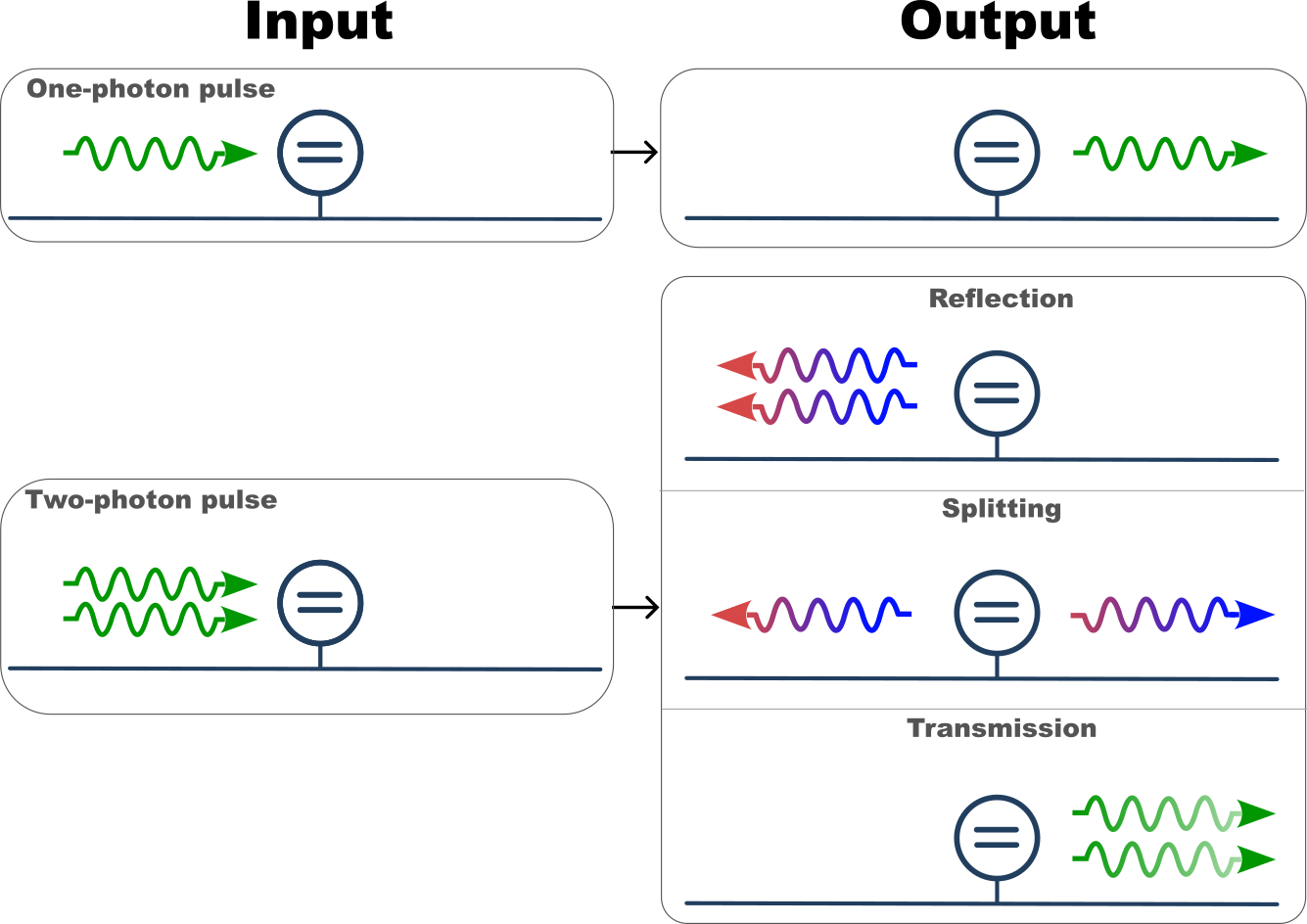}
    \caption{
    Photon scattering on a quadratically-coupled two-level emitter. The nonlinearity of the light-matter interaction is such that the emitter is completely transparent to one-photon pulses, while it strongly interacts with the multi-photon components of the input state. When a two-photon state is sent as input there are three allowed output channels: reflection, splitting, and transmission. The frequency distribution of the input photons plays a key role in determining each channel probability. The frequency of reflected and split photons are strongly anti-correlated, while the transmitted photons are positively correlated.
    }\label{drawing}
  \end{figure}

In this work, we develop a quantum optics theory to describe a single quantum emitter interacting quadratically with the photons that \emph{propagate} along a one-dimensional waveguide. We study this problem with a two-photon scattering theory based on a Wigner-Weisskopf approach and the Born-Markov approximation. We derive the general form of the scattering matrix, including semi-analytical solutions for arbitrary photonic input states and full analytical solutions for Gaussian inputs. Applying this theory, we unveil observable features of the emitter's response that are fundamentally different with respect to conventional waveguide QED setups (see Fig.~\ref{drawing}). These include (i) the spontaneous emission of correlated biphoton states, (ii) the strong interaction with spectrally narrow two-photon pulses, and (iii) full transparency to single-photon inputs. Finally, we show that these effects can be exploited in quantum information applications, designing a deterministic controlled-phase gate that acts on pairs of propagating photons with perfect fidelity. This result seems to contradict a famous no-go theorem for photonic gates~\cite{Shapiro06,GeaBanacloche10,nogo_theorem}. However, it is the use of quadratic interactions that allows us to eliminate the wavepacket distortions that are intrinsic to the localized non-linearities induced by linear light-matter coupling~\cite{nogo_theorem}. The proposed gate is based on dual-rail encoding, which is of increasing relevance for superconducting quantum-computing applications~\cite{teoh22}. All results discussed in this work can be implemented using state-of-the-art superconducting quantum emitters~\cite{felicetti_two-photon_2018,PhysRevA.98.053859,bertet2005dephasing,Steele_sidebands} interacting with propagating microwave fields or nanomechanical oscillators~\cite{Zhou2006} and we expect that similar regimes will soon be achievable with solid-state~\cite{Wang2016,Munoz2018} or atom-based~\cite{felicetti_spectral_2015,PhysRevA.97.023624,PhysRevA.99.032303,Cong_selective,Schneeweiss_2018, Dareau_2018} nanophotonic devices. 

The work is structured as follows. In Sec.~\ref{SII}, we develop a general theory for the scattering of pairs of photons by a single quantum emitter interacting quadratically with a one-dimensional waveguide. In Sec.~\ref{SIII}, we derive consequences from this theory, such as the two-photon spontaneous emission law and the two-photon scattering probabilities. We derive general bounds for the scattering cross-section valid for arbitrary photon-pair (biphoton) input states. We then consider specific examples for spontaneous emission and for the scattering of biphoton states with Gaussian and Lorentzian frequency distributions. In Sec.~\ref{Control-Phase_Gate}, we provide a striking example of application in quantum-information tasks, showing how to design a controlled-phase gate with unit fidelity. In Sec.~\ref{discussion}, we discuss possible physical implementations of the considered model. In Sec.~\ref{Conclusions}, we provide conclusions and the outlook for future directions.  

\section{Scattering theory for two-photon interactions}
\label{SII}
This section introduces a new scattering theory for two-photon states impinging on a quadratically-coupled quantum emitter. It describes the model's Hamiltonian \eqref{1_model}, the derivation of the scattering equations using the Wigner-Weisskopf formalism \eqref{2_WW}, and the computation of the two-photon scattering matrix under the Markov approximation \eqref{sectionscatteringtheory}.

\subsection{Model}
\label{1_model}

We consider a two-level quantum emitter with general quadratic coupling to a continuum of bosonic modes in a one-dimensional waveguide. Under the rotating-wave approximation, the system Hamiltonian can be written as ($\hbar = 1$),
\begin{align}\label{Htwophoton}
\hat{H}&=\omega_0\hat{\sigma}_+\hat{\sigma}_-+\sum_{\mu\in\{\pm\}}^{}\int_{}^{}\omega {\hat{a}_\omega^{\mu\dagger}}\hat{a}_\omega^\mu\,d\omega \nonumber\\
    \quad &+\sum_{\mu,\mu'\in \{\pm\}}\int g^{\mu\mu'}_{\omega{\omega'}}\hat{\sigma}_+\frac{\hat{a}_\omega^\mu\hat{a}_{\omega'}^{\mu'}}{\sqrt{2}}\,d\omega d\omega'+(H.c.).
\end{align}
Here, $\omega_0$ and $\hat{\sigma}_+$ ($\hat{\sigma}_-$) are the resonance frequency and the raising (lowering) operator of the emitter, respectively. The operator ${\hat{a}_\omega}^\mu$ (${\hat{a}_\omega^{\mu\dagger}}$) annihilates (creates) a waveguide photon with frequency $\omega$ and propagating direction $\mu\in\{\pm\}$.
We denote with $g^{\mu\mu'}_{\omega{\omega'}}$ the function that gives the quadratic coupling strength between the emitter and the field frequency components.
Note that  Eq.~\eqref{Htwophoton} is over-parametrized. This is because the relation 
$[\hat a_\omega^\mu,(\hat a_{\omega'}^{\mu'})^\dagger]=\delta_{\mu,\mu'}\delta_{\omega,\omega'}$
implies that all vector elements appear twice in the summation and integration. Therefore, we can arbitrarily set the symmetric condition $g^{\mu\mu'}_{\omega{\omega'}}=g^{\mu'\mu}_{\omega'{\omega}}$, holding for any $\mu$, $\mu'$, $\omega$ and $\omega'$.
We will keep the coupling function as general as possible in the derivation, specifying it only in the examples, in order to identify the intrinsic properties of the quadratic coupling. The scattering theory that we develop is applicable to different experimental implementations of the model.

\subsection{Wigner-Weisskopf ansatz}
\label{2_WW}

The Hamiltonian in Eq.~\eqref{Htwophoton} commutes with the operator $\hat{N} = \sum_\mu \int{\hat{a}_\omega^{\mu\dagger}}\hat{a}_\omega^\mu d\omega + 2\hat{\sigma}_+ \hat{\sigma}_-$. This is a continuous-symmetry generator associated with the conservation of the weighted number of excitations~\cite{felicetti_spectral_2015}. This conservation law implies that the whole dynamics can be described using a Wigner-Weisskopf ansatz wavefunction with a fixed number of (weighted) excitations as defined by $\hat{N}$. In the vacuum sector or in the sector with individual photons, this dynamics is trivial: in both cases, the photonic states are decoupled from the emitter, which is transparent to individual photon states.

Consequently, we may focus on solving the scattering problem for input states with either two incoming photons and an emitter in the ground state, or an excited emitter in the vacuum. The most general such state,
\begin{align}\label{state}
\ket{\Phi(t)}&=C_e(t)\hat{\sigma}_+ \ket{\textbf{0}}\nonumber\\
\quad&+\sum_{\mu,\mu'\in \{\pm\}}\int C^{\mu\mu'}_{\omega{\omega'}}(t)\frac{(\hat{a}_\omega^\mu \hat{a}_{\omega'}^{\mu'})^\dagger}{\sqrt{2}}\,d\omega d\omega'    \ket{\textbf{0}},
\end{align}
is constructed by creating excitations on top of \(\ket{\textbf0}\), the tensor product between the bosonic vacuum and the emitter's ground state. The two-photon Wigner-Weisskopf state is normalized as
\begin{equation}
    \left|C_e(t)\right|^2+\sum_{\mu,\mu'\in \{\pm\}}\int \left|C^{\mu\mu'}_{\omega{\omega'}}(t)\right|^2\,d\omega d\omega'=1,
\end{equation}
and it holds that $C^{\mu\mu'}_{\omega{\omega'}}(t)=C^{\mu'\mu}_{\omega'{\omega}}(t)$ for any $\mu$, $\mu'$, $\omega$, $\omega'$.

The Wigner-Weisskopf ansatz~\eqref{state} describes a solution to the Schrödinger equation,
$i\partial_t\ket{\Phi(t)}=\hat{H}\ket{\Phi(t)}$, provided the coefficients satisfy a set of coupled, linear ordinary differential equations:
\begin{align}
\label{system1}
    &i\Dot{C_e}(t) = \omega_0C_e(t)+\sum_{\mu,\mu'\in\{\pm\}}\int{ {g^{\mu\mu'}_{\omega{\omega'}}} C^{\mu\mu'}_{\omega{\omega'}}(t)d\omega d\omega'}, \\
    &i\Dot{C}^{\mu\mu'}_{\omega{\omega'}}(t) = (\omega+\omega')\label{system2} C^{\mu\mu'}_{\omega{\omega'}}(t)+(g^{\mu\mu'}_{\omega{\omega'}})^\ast C_e(t).
\end{align}
It is convenient to reparameterize the wavefunction's element using the sum and difference of photon frequencies, $\overbar{\omega}=\omega'+\omega$ and   $\Delta=\omega'-\omega$, as $C_{\overbar{\omega}\Delta}^{\mu\mu'}(t)$ and 
$g^{\mu\mu'}_{\overbar \omega \Delta}$.

At this stage, let us introduce the Markovian approximation, assuming that the emitter will spontaneously decay, releasing photons to the waveguide at a rate that is much slower than its intrinsic frequency, $\Gamma \ll \omega_0$. In this limit one can formally solve Eq.~\eqref{system2} and replace the solution into Eq.~\eqref{system1} to obtain (see App.~\ref{appendix_self}): 
\begin{align}\label{Cdot}
    &i\Dot{C_e}(t)=-i\left(\frac{\Gamma}{2}+i\omega_0\right)C_e(t) + \nonumber\\
    &\sum_{\mu,\mu'\in\{\pm\}}
       \int_0^\infty d\overbar\omega \int_{0}^{\overbar\omega}d\Delta
    g^{\mu\mu'}_{\overbar\omega\Delta}
    {C}^{\mu\mu'}_{\overbar{\omega}{\Delta}}(t_0)e^{-i\overbar{\omega}(t-t_0)}.
\end{align}
The total spontaneous emission rate is self-consistently defined as
\begin{equation}\label{Gammagrande}
    \Gamma=\sum_{\mu,\mu'\in\{\pm\}} \int_{0}^{\omega_0}
2\pi\, |g^{\mu\mu'}_{\omega_0 \Delta}|^2 \, d\Delta \; \ll \omega_0.
\end{equation}

In the standard Markovian approach with linear interactions, the decay rate depends on the frequency of the emitted photons, although this dependency is usually weak or ignored~\cite{QuantumNoise}. In the case of quadratic light-matter interactions, the decay rate depends on the coupling strength evaluated at the two-photon resonance, {\it i.e.} at $\overbar \omega = \omega_1 + \omega_2 = \omega_0$, and integrated over frequency differences $\Delta$. We can assume that the coupling strength does not change significantly with $\overbar{\omega}$, at least in a band of frequencies around $\omega_0$, but we cannot fully eliminate the dependency on $\Delta$, thus $g_{\omega\omega'}^{\mu\mu'}\equiv g_{\Delta}^{\mu\mu'}$. This dependency may be factored into a product of emission rates $\gamma^{\mu\mu'}\in \mathbb{R}$ and an envelope~\(u(\Delta)=u(-\Delta)\in\mathbb{C}\), i.e.
\begin{equation}\label{newgamma}
g^{\mu\mu'}_{\Delta}=\sqrt{\frac{\gamma^{\mu\mu'}}{2\pi}}u(\Delta).
\end{equation}
We consider $\omega_0$ to be the dominant energy scale, that is we take $\omega_0\rightarrow\infty$ for the upper integration limit in Eq.~\eqref{Gammagrande}. This corresponds to assuming that the coupling function quickly decays with $\Delta$. Then, we take $u(\Delta)$ to be a bell-shaped function normalized to 1 in the $l^2$-norm, i.e. 
$\|u\|_{l^2}/\sqrt{2}=1$~\footnote{The $l^2$-norm is defined as $\| f\|_{l^2}=\sqrt{\int |f(x)|^2dx}$}. This convenient normalization implies that the function $u(\Delta)$ encodes the difference-frequency dependence of the coupling strength, but it does not affect the total emission rate $\Gamma$. As we will see in the following, the explicit dependence on $\Delta$ has highly non-trivial consequences on the scattering phenomenology and on potential applications.

\begin{figure*}[ht!]
    \centering
    \includegraphics[width=1\textwidth]{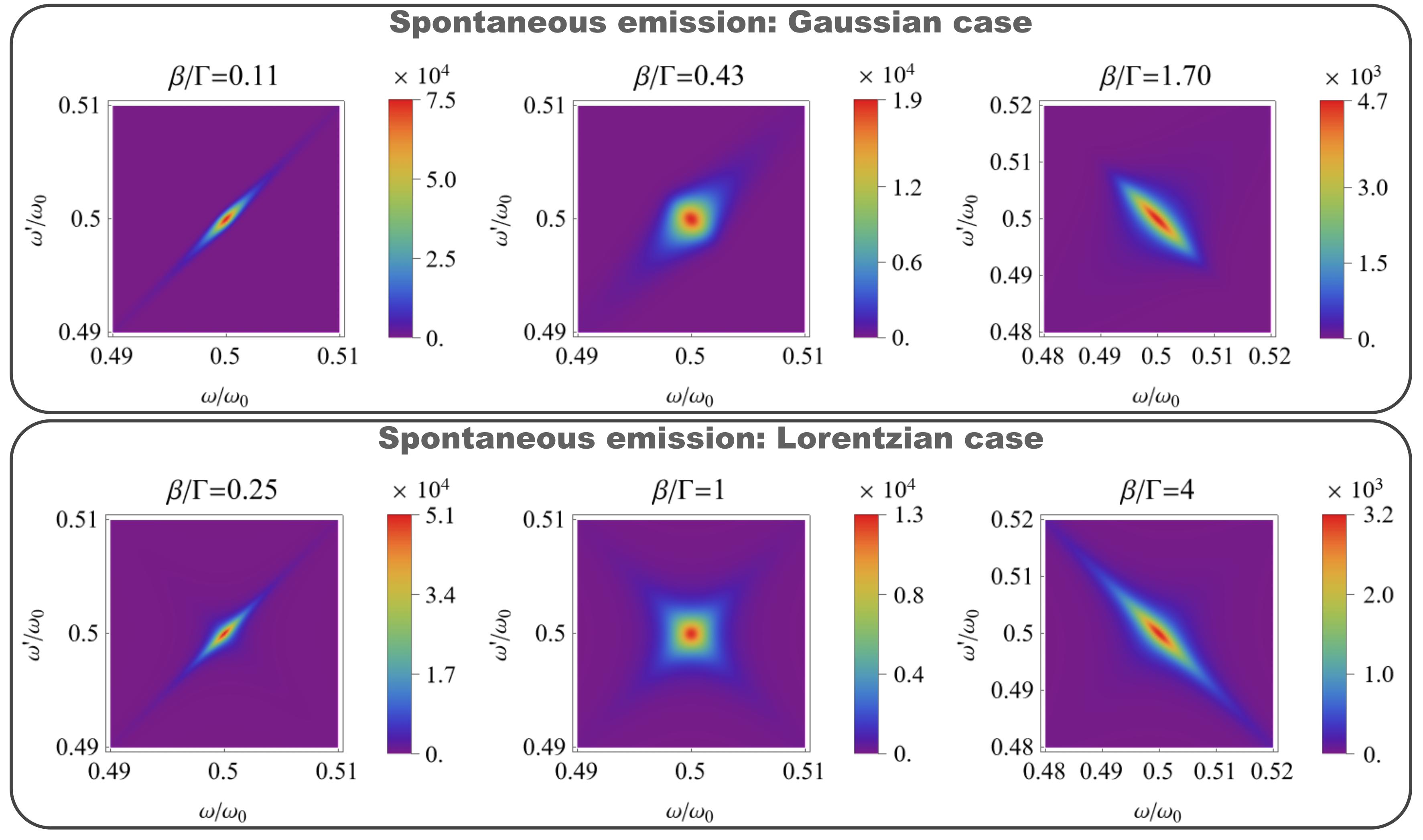}
    \caption{Spontaneous decay: The plots show the frequency distribution (FD) of the emitted field  as a function of the two output frequencies \(\omega\), \(\omega'\). We compare the output FDs for two different coupling functions \(g^{\mu\mu'}_\Delta= u(\Delta)\sqrt{\gamma^{\mu\mu'}/2\pi}\), with an isotropic spontaneous emission rate \(\gamma^{\mu\mu^\prime}=\Gamma/4\), where \(\Gamma=0.004\omega_0\). In the first line, we consider the Gaussian function of Eq.~\eqref{couplinggaussiano}, while in the second line we assume the Lorentzian coupling of Eq.~\eqref{Lorentzg}. The parameters $\beta$ of the coupling functions are chosen to have the same FWHM in both cases. Notice how the spectrum of the spontaneously emitted photons switches from frequency correlation to anticorrelation as the ratio \(\beta/\Gamma\) increases.  An analysis of the entanglement properties of spontaneously-emitted photon pairs is provided in Sec.~\ref{Sec_Entanglement}.
    }
    \label{spontaneous_FD}
    \end{figure*}

\subsection{Two-photon scattering matrix}\label{sectionscatteringtheory}

The scattering matrix connects linearly the amplitude coefficient of the input photons ${C}^{\mu\mu'}_{\overbar{\omega}\Delta}(t_0)$ with the output field ${C}^{\mu\mu'}_{\overbar{\omega}\Delta}(t_1)$ at asymptotic times $t_1\gg t_0$, when all the interaction with the quantum emitter has ceased. As explained in App.~\ref{appendix_scattering}, it is derived by integrating Eq.~\eqref{system2} using the initial and the final times as boundary conditions and equating the results. We obtain:
\begin{equation}
\label{InOut}
C^{\mu\mu'}_{\overbar \omega \Delta}(t_1) e^{i\overbar \omega t_1} =e^{i\overbar\omega t_0} C^{\mu\mu'}_{\overbar \omega \Delta}(t_0) +
i\sqrt{\gamma^{\mu \mu'}} u(\Delta)^* \tilde{C}_e(\overbar \omega),
\end{equation}
where we defined $\tilde{C}_e(\omega) = \frac{1}{\sqrt{2\pi}}
\int_{t_0}^{t_1} C_e(\tau) e^{i\omega \tau}$. This quantity may be obtained from the Fourier transform of Eq.~\eqref{Cdot}, i.e.
\begin{equation}\label{Fourier_C_e}
\tilde{C}_e(\overbar \omega) = -\sum_{\mu \mu'}\frac{i \sqrt{\gamma^{\mu \mu'}} e^{i\overbar\omega t_0 }}{\frac{\Gamma}{2} + i (\omega_0 - \overbar\omega)} \int_0^{\overbar\omega} u(\Delta) C_{\overbar \omega \Delta}^{\mu \mu'}(t_0) d\Delta,
\end{equation}
assuming that the emitter is initially in the ground state $C_e(t_0)=0$ and that there is a large temporal separation between the input and the output events. More precisely, we require that $(t_1-t_0)\gg 1/\Gamma$, so that the emitter is completely decayed at the time in which the output state is formally defined $C_e(t_1)=0$. For simplicity, in the following we consider the narrow-bandwidth limit and extend the integration region to $\infty$ in Eq.~\eqref{Fourier_C_e} (see App.~\ref{appendix_scattering}). This assumption is not strictly needed to obtain semi-analytical results.

For the sake of clarity, let us first write the scattering relations for a fixed propagation direction of the input field along the waveguide. In other words,  we set $C_{\overbar{\omega}\Delta}^{\mu\mu'}(t_0)=C_{\overbar{\omega}\Delta}(t_0)\delta^{\mu}_{\lambda}\delta^{\mu'}_{\lambda'}$ for some fixed \(\lambda\) and \(\lambda'\), where \(\delta^{\mu}_\lambda\) is a Kronecker delta. In this way, we can remove the summation in the direction index in Eq.~\eqref{Fourier_C_e}. This assumption will be dropped in the final expression of the scattering matrix. After all these manipulations, the output and input states are related through
\begin{align}
&C^{\mu\mu'}_{\overbar \omega \Delta}(t_1)=  e^{-\overbar \omega(t_1 - t_0)} 
\left\{ C_{\overbar\omega \Delta}(t_0)
\delta_{\lambda}^\mu \delta_{\lambda'}^{\mu'}+ \right.\nonumber \\
& \left. 
-\frac{\sqrt{\gamma^{\mu \mu'} \gamma^{\lambda \lambda'}} }{\frac{\Gamma}{2} + i (\omega_0 - \overbar\omega)} u(\Delta)^* \int_0^{\infty} d\Delta 
u(\Delta) C_{\overbar \omega \Delta}(t_0)
\right\}.
\end{align}
As in the standard case of dipolar interactions, the output field is the interference between the input biphoton state (first line) and the two photons re-emitted by the atom (second line). Also as in standard waveguide QED, the Markovian approximation manifests itself in the Lorentzian dependency with respect to the total energy of the input field $\overbar \omega$. However, differently from the dipolar case, for quadratic interactions the re-emitted field depends on the relative frequency $\Delta$ through the interaction envelope $u(\Delta)$. In the next sections, we will explore the consequences of this relationship. Let us now write a general relation for the two-photon scattering matrix.
\newpage
\begin{prop}\label{prop1}{\bf [Scattering matrix]}

Consider a quadratic interaction between bosonic modes and a two-level emitter, defined by the Hamiltonian in Eq.~\eqref{Htwophoton}, under Markovian approximation $\Gamma \ll \omega_0$. Given a two-photon input state $|\Phi(t_0)\rangle$ as in Eq.~\eqref{state}, the probability amplitudes at a time $t_1$ such that $(t_1-t_0)\Gamma\gg 1$ are  
\begin{equation}\label{scatteringinputoutput}
{C}^{\mu\mu'}_{\overbar{\omega}\Delta}(t_1)=e^{-i\overbar{\omega}(t_1-t_0)}\int_{0}^{\infty}d\Delta' S^{\mu\mu'}_{\alpha\alpha'}(\overbar{\omega}, \Delta,\Delta'){C}^{\alpha\alpha'}_{\overbar{\omega}\Delta'}(t_0),
\end{equation}
where a summation is implicit over the repeated indexes $\alpha$ and $\alpha'$. The term $S^{\mu\mu'}_{\alpha\alpha'}(\overbar{\omega}, \Delta,\Delta')$ is a generic element of the scattering matrix, which can be expressed as
\begin{align}\label{scatteringmatrix}
S^{\mu\mu'}_{\alpha\alpha'}(\overbar{\omega}, \Delta,\Delta') =
\delta^{\mu}_\alpha\delta^{\mu'}_{\alpha'}\delta(\Delta-\Delta')  
+u(\Delta)^\ast u(\Delta') \Theta_{\alpha \alpha'}^{\mu \mu'}(\overbar \omega),
\end{align}
where we defined 
\begin{equation}
\Theta_{\alpha \alpha'}^{\mu \mu'}(\overbar \omega) = 
- \frac{\sqrt{\gamma^{\alpha \alpha'}\gamma^{\mu \mu'}  }}{ \frac{\Gamma}{2} + i (\omega_0 -\overbar \omega ) }.
\end{equation}
\end{prop}

In Eqs.~\eqref{scatteringinputoutput} and~\eqref{scatteringmatrix}, we identify that the integration with $u(\Delta)u(\Delta')$ plays the role of a projection onto a specific biphoton distribution $u(\Delta)$. The action of the scattering matrix onto the input state is better understood if we explicitly separate the parallel and orthogonal components to $u(\Delta)$ as
\begin{equation}\label{decomposition}
    {C}^{\mu\mu'}_{\overbar{\omega}\Delta}(t) = 
    {C}^{\mu\mu'}_{\overbar{\omega}\Delta,\bot}(t) +
    {C}^{\mu\mu'}_{\overbar{\omega}\Delta,\|}(t),
\end{equation}
with
\begin{equation}\label{C_parallel}
{C}^{\mu\mu'}_{\overbar{\omega}\Delta, \|}(t) = 
u(\Delta)^\ast \int_{0}^\infty d\Delta'
u(\Delta'){C}^{\mu\mu'}_{\overbar{\omega}{\Delta}}(t).
\end{equation}
The scattering relation can then be rewritten as
\begin{equation}\label{Scattering_parOrth}
{C}^{\mu\mu'}_{\overbar{\omega}\Delta}(t_1) = 
e^{-i\overbar \omega (t_1 - t_0)}
\left[ \chi_{\alpha \alpha'}^{\mu \mu'} (\overbar \omega)\ 
C^{\alpha \alpha'}_{\overbar\omega \Delta,\|}(t_0)
+ {C}^{\mu\mu'}_{\overbar{\omega}\Delta,\bot}(t_0) \right].
\end{equation}
The orthogonal component ${C}^{\mu\mu'}_{\overbar{\omega}\Delta,\bot}$ is unperturbed by the quantum emitter. The parallel component experiences a transformation $\chi_{\alpha \alpha'}^{\mu \mu'} (\overbar \omega) = 
\delta_{\alpha}^{\mu}\delta_{\alpha'}^{\mu'} + \Theta_{\alpha \alpha'}^{\mu \mu'}(\overbar \omega) $ which is analogous of the scattering phase for a single photon in waveguide QED with linear Markovian interaction~\cite{fan_input-output_2010}.

\section{Paradigmatic cases}
\label{SIII}
In this section, we study the phenomenology of spontaneous emission \eqref{Spontaneous_emission} and  two-photon scattering \eqref{Two-photon_Scattering}, analyzing the cross-section and the spectral features of the output channels. In the case of scattering of biphoton states, we provide general bounds on the scattering probabilities which hold for arbitrary frequency distributions of the input. We then focus on the Gaussian case, where we are able to give explicit analytical solutions for the scattering matrix and gather physical intuition on the features of the scattering problem. We also reproduce the analysis in the semi-analytical case of Lorentzian coupling function, which confirms that the results of the Gaussian case are valid for biphoton states of high experimental relevance.

\subsection{Spontaneous emission}
\label{Spontaneous_emission}

When we excite the quantum emitter in a waveguide without photons and both interact via Eq.~\eqref{Htwophoton}, we expect the emitter to relax to the ground state emitting two photons. We can use the Wigner-Weisskopf theory [see Eq.~\eqref{Cdot}] to derive the frequency distribution of the emitted photons at asymptotic times. In the Markovian regime, the amplitude of the emitter's excited state decays exponentially with the rate $\Gamma$ introduced in Eq.~\eqref{Gammagrande}:
\begin{equation}\label{Cesol}
C_e(t)=e^{-i\omega_0(t-t_0)}e^{-\frac{\Gamma}{2}(t-t_0)}.
\end{equation}
The emitted photon wavepacket in frequency space at long times $t_1\gg t_0$ is recovered by inserting Eq.~\eqref{Cesol} into Eq.~\eqref{InOut}. This wavefunction,
\begin{align}\label{Comegaspontaneous}
\abs{{C}^{\mu\mu'}_{\overbar{\omega}{\Delta}}(t_1)}^2=
 \frac{1}{2\pi}\frac{\gamma^{\mu\mu'} |u(\Delta)|^2}{\Gamma^2/4+(\omega_0-\overbar{\omega})^2}\,,
\end{align}
depends on the sum $\overbar{\omega}$ and on the difference of frequencies $\Delta$. This profile resembles the Lorentzian lineshape that is characteristic of Markovian dynamics, where the wavepacket is centered around the total energy of $\overbar{\omega}$ of the incoming photons. However, the actual wavepacket is now shaped by the dependence of the interaction on the relative frequency of the photons $g_{\Delta}^{\mu\mu'}$. This is a general result that can be applied to very different experimental scenarios.

{\it Example: Gaussian and Lorentzian coupling functions--}
As an example, let us compute the spectrum of the spontaneously emitted photon pairs in Eq.~\eqref{Comegaspontaneous}, assuming two different frequency distributions for the coupling function. The first coupling function is Gaussian with zero mean and variance \(\beta^2\), i.e.
\begin{equation}\label{couplinggaussiano}
g^{\mu\mu',{\rm G}}_\Delta= \sqrt{\frac{\gamma^{\mu\mu'}}{2\pi}} \left(\frac{2}{\pi\beta^2}e^{-\frac{\Delta^2}{\beta^2}}\right)^{1/4}.
\end{equation}
For comparison, we consider also a Lorentzian distributed coupling
\begin{equation}
\label{Lorentzg}
g^{\mu\mu',{\rm L}}_\Delta= 
\sqrt{\frac{\gamma^{\mu\mu'}}{2\pi}}
\sqrt{\frac{1}{\pi} \left( \frac{\beta}{\beta^2/4 + \Delta^2}\right) }.
\end{equation}
 This gives us an expression for $|C_{\overbar \omega \Delta}^{\mu\mu'}(t_1)|^2$ via Eq.~\eqref{Comegaspontaneous}.  In Fig.~\ref{spontaneous_FD}, we plot the emission rate for three different values of \(\beta\), in the isotropic case where \(\gamma^{\mu\mu^\prime}=\Gamma/4\). Notice that, unless a non-isotropic emitter or a chiral waveguide is considered~\cite{lodahl_chiral_2017}, 
there are three possible output channels, such that the photon pairs are emitted on the same direction or they can split. The frequency distribution is shown in Fig.~\ref{spontaneous_FD} and it is the same for the three possible channels.

Summarizing, similarly to standard linear couplings the decay of the emitter excitation is exponential. Strong elements of novelty are: (i) Spontaneous decay of a single emitter results in the generation of a photon pair. (ii) The emitted biphoton state has a non-trivial frequency distribution, which is Lorentzian in the sum-frequency variable (main diagonal in Fig.~\ref{spontaneous_FD}, width set by the total decay rate $\Gamma$), while it reproduces the coupling function in the difference-frequency variable [anti-diagonal in Fig.~\ref{spontaneous_FD}, width set by the full width at half maximum, FWHM, of $u(\Delta)$]. (3) Interestingly, it is possible to generate frequency-correlated or anticorrelated photon pairs by changing the ratio between the total emission rate $\Gamma$ and the FWHM of the coupling function $u(\Delta)$.

\subsection{Scattering of two-photon inputs}
\label{Two-photon_Scattering}

\begin{figure*}[ht!]
    \centering
    \includegraphics[width=0.90\textwidth]{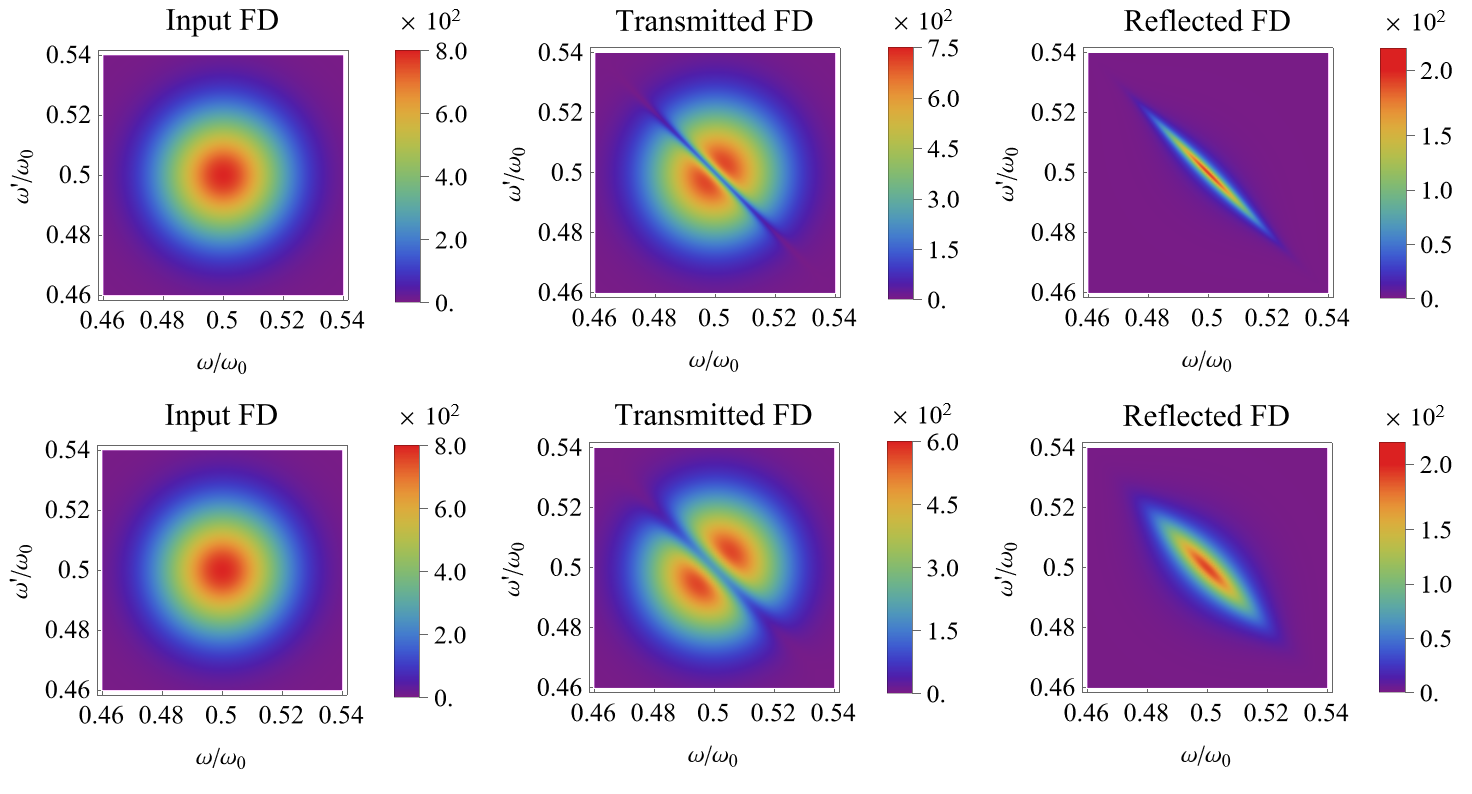}
    \caption{Two photon scattering, Gaussian case. We compare the transmitted and reflected/splitted frequency distributions (FDs) for two different values of the total emission rate, \(\Gamma = 0.004\omega_0\) for the first and \(\Gamma=0.012\omega_0\) for the second line. The FD for the reflection and splitting processes are identical. The FDs are shown in function of the two frequency variables $\omega$, $\omega'$. The input field has been chosen as the product of two independent Gaussian FDs in \(\overbar{\omega}\) and \(\Delta\), centered at \(\omega_0\) (corresponding to the two-photon resonance condition), variance \(\alpha^2=(0.02\omega_0)^2\) and FWHM\(=2\alpha\sqrt{2\ln2}=0.047\omega_0\).  We consider a Gaussian distributed coupling [Eq.~\eqref{couplinggaussiano}] with \(\beta=\alpha\) and we assume an isotropic spontaneous emission rate, i.e. \(\gamma^{\mu\mu^\prime}=\Gamma/4\).  Focusing on the scattered field (rightmost plots), we see that the width in the sum-frequency variable $\overbar{\omega}$ (main diagonal) is proportional to \(\Gamma\), while the width in the difference-frequency variable $\Delta$ (anti-diagonal) is given by the input field variance $\alpha^2$ (which has been set equal to that of the coupling function). Interestingly, we see that, when both photons are transmitted, their frequency distribution is correlated, while it is strongly anti-correlated for the reflection and splitting processes.}
    \label{outputFD}
\end{figure*}

Let us discuss the phenomenology that appears in the scattering dynamics computed in Sec.~\ref{sectionscatteringtheory}. As sketched in Fig.~\ref{drawing}, for two copropagating input photons $\lambda\lambda'=++$, there are three possible outcomes classified by the output directions: (i) reflection ($\mu\mu'=--$), (ii) transmission ($\mu\mu'=++$) and (iii) splitting ($\mu\neq\mu'$), depending on whether both directions are preserved or reversed, or become opposite to each other. The respective scattering probabilities integrated over the complete frequency range are
\begin{equation}\label{coefficienti1}P^{\mu\mu'}=\frac{\iint |C^{\mu\mu'}_{\overbar{\omega}\Delta}(t_1)|^2d\Delta d\overbar{\omega}}{\iint |C^{++}_{\overbar{\omega}\Delta}(t_0)|^2d\Delta d\overbar{\omega}}.
\end{equation}
These values depend on the particular input field, but they exhibit general symmetries $P^{-+}=P^{+-}$ and constraints that can be derived from Prop.~\ref{prop1}.

Let us find an upper bound to the reflection probability under reasonable assumptions on the input state. This corresponds to maximizing Eq.~\eqref{coefficienti1} for \(\mu\mu'=--\). 
Let us assume the atom to be initially in its ground state and the input field probability amplitude to be separable in $\Delta$ and $\overbar{\omega}$, i.e. 
\begin{equation}
C^{++}_{\overbar{\omega}\Delta}(t_0)=f(\overbar{\omega})h(\Delta),
\end{equation}
where $f(\overbar{\omega})$ and $h(\Delta)$ are complex functions normalized in the $l^2$-norm, i.e. $\|f\|_{l^2}=1=\|h\|_{l^2}/\sqrt{2}$.
By using the Cauchy-Schwarz inequality in Eq.~\eqref{scatteringinputoutput}, we obtain 
\begin{align}\label{Massimizzare}
    \left|{C}^{--}_{\overbar{\omega}\Delta}(t_1)\right|\leq \left|{C}^{++}_{\overbar{\omega}\Delta}(t_0)\right|\cdot\left|\Theta^{--}_{++} (\overbar{\omega})\right|.
\end{align}
The inequality in Eq.~\eqref{Massimizzare} is saturated if and only if  \(h(\Delta)=e^{i\varphi}u(\Delta)^\ast\), that is when the frequency distribution of the input field matches the coupling function $g^{\mu\mu'}_{\Delta}=g^{\mu\mu'}u(\Delta)$.
Inserting Eq.~\eqref{Massimizzare} into Eq.~\eqref{coefficienti1} and using Holder's inequality, we obtain that the reflection probability $R\equiv P^{--}$ is bounded by
\begin{align}\label{boundP}
    R&\leq \frac{1}{2}\| f\|^2_{l^2}\|h\|^2_{l^2}\|\Theta^{--}_{++}\|^2_{l^\infty}=\frac{4\gamma^{++}\gamma^{--}}{\Gamma^2}\equiv R_{\rm max}. 
\end{align}
This inequality is saturated when the bandwidth $\delta$ of the input field $f(\overbar{\omega})$---measured as the FWHM or by some other means---is much narrower than the emitter's natural linewidth $\Gamma\gg\delta$.

We can also derive exact expressions for the maximal achievable rates of splitting $S\equiv 2P^{-+}$ and transmission $T\equiv P^{++}$ processes, defined as
\begin{align}
S&\leq \frac{8\gamma^{++}\gamma^{-+}}{\Gamma^2}\equiv S_{\rm max}, \\
T&\geq 1-R_{\rm max}-S_{\rm max}\equiv T_{\rm min}.
\end{align}
While in waveguide QED with linear interactions a single emitter can perfectly reflect a single photon~\cite{hoi_demonstration_2011}, the perfect two-photon reflection by a quadratically-coupled emitter is only possible when the splitting process is suppressed, i.e.  $\gamma^{-+}=0$ and $\gamma^{++}=\gamma^{--}$.
In the case of isotropic interactions $\gamma^{\mu\mu^\prime}=\Gamma/4$, this is thus not possible. Here, even for optimal frequency-matched input \(h(\Delta)= e^{i\varphi}u(\Delta)^\ast\), one obtains $R_{\rm max}=1/4$, $S_{\rm max}=1/2$ and $T_{\rm min}=1/4$.

\begin{figure}[ht!]
\center
\includegraphics[width=.87\linewidth]{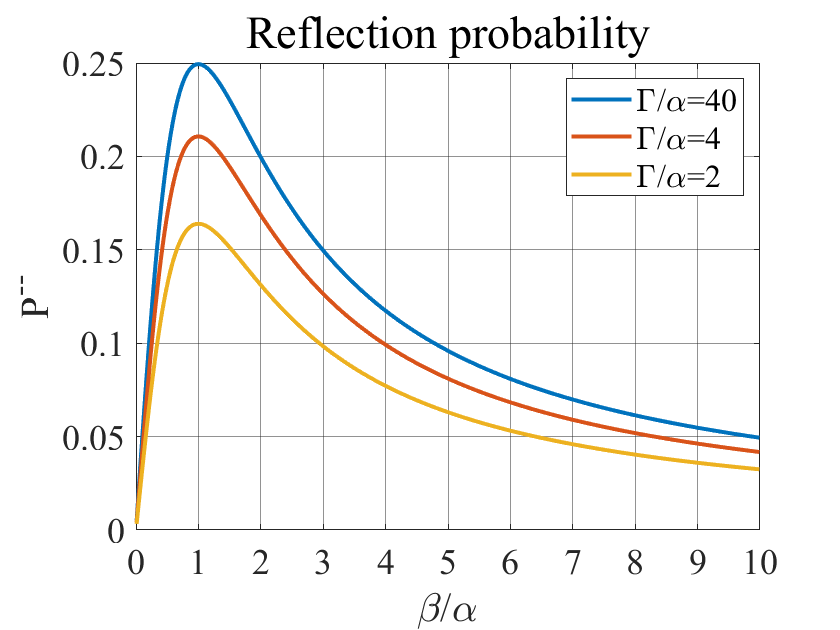}
    \caption{Plot of the reflection probability, i.e. \(P^{--}\), as a function of the ratio \(\beta/\alpha\). To obtain the plot, we considered a Gaussian input [Eq.~\eqref{Gauss_prod}] centered in \(\omega_0\) with variance \(\alpha^2\) and a Gaussian coupling with variance \(\beta^2\) [Eq.~\eqref{couplinggaussiano}]. In addition, we assumed an isotropic spontaneous emission rate. We compare the reflection probability for different values of \(\Gamma\) and we show that the maximum value of \(P^{--}\) is reached when \(\alpha=\beta\) and \(\Gamma\gg\alpha\). }\label{graficoriflessione}
  \end{figure}

{\it Example: Gaussian wavepackets--}
Let us discuss a scattering experiment with Gaussian couplings and Gaussian input wavepackets that has analytical solutions and which provides insight that is extensible to other situations, e.g., the  Lorentzian case shown in App.~\ref{appendix_Lorentzian}. Without loss of generality, we use the isotropic Gaussian coupling function with variance $\beta^2$ from Eq.~\eqref{couplinggaussiano}. The input state is formed by two Gaussian wavepackets with frequencies $\omega_{1,2}$, propagating along the same direction $(++)$ with identical widths \(\alpha^2\). The state is separable both on the individual photons' frequencies and the canonical variables $\{\overbar{\omega},\Delta\}$
\begin{equation}
\label{Gauss_prod}
C^{\mu\mu'}_{\overbar{\omega}\Delta}(t_0)=\frac{\delta^\mu_+\delta^{\mu'}_+}{\alpha\sqrt{\pi}}e^{-\frac{[\overbar{\omega}-(\omega_1+\omega_2)]^2}{4\alpha^2}}e^{-\frac{[\Delta-(\omega_1-\omega_2)]^2}{4\alpha^2}}.
\end{equation}
We can now use the scattering matrix formalism to retrieve the expression of the output amplitude probabilities (see Prop.~\ref{prop1}).  Indeed, using Eq.~\eqref{scatteringinputoutput} and Eq.~\eqref{couplinggaussiano}, we find the frequency distribution of the  field for the different output channels
\begin{align}\label{newComegascattering}
  &C^{\mu\mu'}_{\overbar{\omega}\Delta}(t_1)=e^{-i\overbar{\omega}(t_1-t_0)}\Big[C^{++}_{\overbar{\omega}\Delta}(t_0)\delta^{\mu}_\lambda\delta^{\mu'}_{\lambda'}\nonumber\\
 &-\sqrt{\frac{1}{8\pi(\alpha^2+\beta^2)}}\frac{\Gamma^{} e^{-\frac{\Delta^2}{4\beta^2}}e^{-\frac{(\omega_1-\omega_2)^2}{4(\alpha^2+\beta^2)}}}{\frac{\Gamma}{4}+i(\omega_0-\overbar{\omega})}e^{-\frac{[\overbar{\omega}-(\omega_1+\omega_2)]^2}{4\alpha^2}}\Big].
\end{align}
Let us now analyze the phenomenology described by this analytical result, focusing on the spectral properties of the output field and the scattering probabilities.

In Fig.~\ref{outputFD}, we plot the spectral distribution of the reflected (${\mu,\mu^\prime} = {-,-}$), split (${\mu,\mu^\prime} = {\pm,\mp}$) and transmitted (${\mu,\mu^\prime} = {+,+}$) states, derived from  Eq.~\eqref{newComegascattering} for two values of \(\Gamma\). The distribution of the scattered/split photons is strongly anti-correlated in frequency, even though the input state was a separable one. A careful observation of the formulas reveals how the output field can be tailored by choosing the input state and by engineering the coupling function. In the sum-frequency variable $\overbar{\omega}$ [main diagonal in Fig.~\ref{outputFD}], the distribution has a Lorentzian shape whose width is established by the intensity of the light-matter coupling strength $\Gamma$. A larger value of the parameter \(\Gamma\) translates into a larger FWHM for the reflected frequency distribution. In the difference-frequency variable $\Delta$ [anti-diagonal in Fig.~\ref{outputFD}], the distribution has a Gaussian shape whose width depends on the widths of the input distribution $\alpha$ and of the coupling function $\beta$.

\begin{figure*}[ht!]
    \centering
\includegraphics[width=0.99\textwidth]{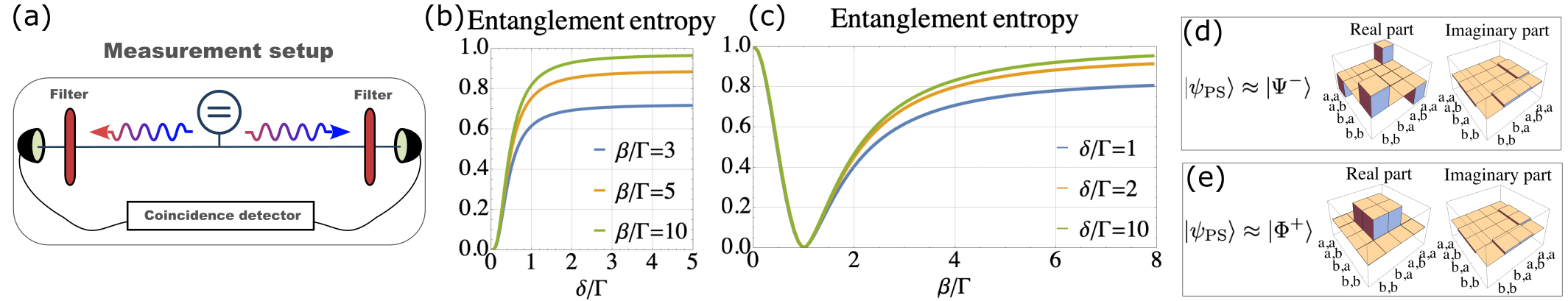}
    \caption{Entanglement generation. (a) Setup of a measurement scheme to post-select spontaneously-emitted photons traveling in opposite directions. (b) Entanglement entropy of the post-selected state of Eq.~\eqref{freqtwostates}, as a function of the detuning of the filters with respect to the two-photon resonance condition. (c) Entanglement entropy as a function of the ratio between the width $\beta$ of the coupling function [see Eq.~\eqref{Lorentzg}] and the total emission rate $\Gamma$. (d) For 
    $\beta/\Gamma = 1/8$ and $\delta/\Gamma = 10$, the state is well approximated by a frequency correlated Bell state $\ket{\psi^-}\equiv 
\frac{1}{2}\left(\ket{\omega_a,\omega_a} - \ket{\omega_b,\omega_b} \right)$. (e) For 
    $\beta/\Gamma = 10$ and $\delta/\Gamma = 10$, the state is well approximated by a frequency anti-correlated Bell state $\ket{\phi^+}\equiv 
\frac{1}{2}\left(\ket{\omega_a,\omega_b} + \ket{\omega_b,\omega_a} \right)$.}
    \label{Entanglement}
\end{figure*}

In Fig.~\ref{graficoriflessione}, we analyze how the total reflection rate depends on the input frequency distribution and the emitter properties. We plot $P^{--}$ as a function of the ratio between the standard deviation $\beta$ of the coupling function [see Eq.~\eqref{couplinggaussiano}] and the standard deviation $\alpha$ of the sum-frequency distribution of the input field [see Eq.~\eqref{Gauss_prod}]. As expected, the scattering rate is maximal when the input frequency distribution matches the coupling function, a specific occurrence of the general results derived in Sec.~\ref{sectionscatteringtheory}. By probing different emission rates $\Gamma$, Fig.~\ref{graficoriflessione} also confirms that the upper bound on the reflection probability $P^{--} = 1/4$ is attainable when the width of the input frequency distribution is small compared to the emitter line width, $\Gamma\gg\alpha$.

\subsection{Entanglement generation}
\label{Sec_Entanglement}
From the phenomenological analysis of Sec.~\ref{Spontaneous_emission} and Sec.~\ref{Two-photon_Scattering}, one can easily infer that two-photon states scattered or spontaneously emitted by a quadratically-coupled emitter have non-trivial entanglement properties. In general, the scattered/emitted states are hyper-entangled, \textit{i.e.} non-classical correlations are present both in the position and in the frequency degrees of freedom. For the sake of brevity, we will focus here on entanglement generated in the (continuous) frequency variable. We consider a specific measurement setting relevant for practical experiments and applications such as quantum key distribution.
Let us take the case of a spontaneously emitted two-photon state, which can be obtained by plugging the result of Eq.~\eqref{Comegaspontaneous} in the definition of the state Eq.~\eqref{state}.  We consider the setup shown in Fig.~\ref{Entanglement}(a), that is we take the post-selected state $\ket{\psi_{\rm PS}}$ in which the photons are emitted in opposite directions, i.e.
\begin{equation}
\ket{\psi_{\rm PS}} = 
\mathcal{N} \int C(\omega,\omega')(\hat{a}_\omega^+ \hat{a}_{\omega'}^{-})^\dagger\,d\omega d\omega'    \ket{\textbf{0}},
\end{equation}
where $\mathcal{N}$ is a renormalization factor and where we have re-expressed the output frequency distribution $C(\omega,\omega') = C^{+-}_{\overbar \omega,\Delta}(t_1)$ as a function of the frequencies of the output photons $\omega$ and $\omega'$. Bipartite entanglement in a continuous variable such as the frequency can be fully characterized with a Schmidt decomposition~\cite{lamata2005dealing, PhysRevA.105.052429}. However, here we consider a simpler approach that is of direct experimental relevance. We identify two discrete frequency modes by applying frequency filters, defined by the functions $f(\omega)$. The effect of the filters is modeled by applying the following transformation to the state distribution:
\begin{equation}
C(\omega, \omega') \rightarrow 
\iint \, d\overbar\omega d\overbar\omega' 
C(\overbar\omega-\omega, \overbar\omega' - \omega') f(\overbar\omega) f(\overbar\omega').
\end{equation}
We choose the filters to have two narrow transmission windows, so that they are well approximated by the sum of two Dirac deltas centered around frequencies $\omega_a$ and $\omega_b$, 
$f(\overbar\omega) =  \delta(\overbar\omega - \omega_a) + \delta(\overbar\omega - \omega_b )$. After the filters, the (post-selected state) can be written as 
\begin{eqnarray}\label{freqtwostates}
\ket{\psi_{\rm PS}} &=& C(\omega_a, \omega_a) \ket{\omega_a, \omega_a} 
+ C(\omega_a, \omega_b) \ket{\omega_a, \omega_b} \nonumber \\ 
&+& C(\omega_b, \omega_a) \ket{\omega_b, \omega_a} 
+ C(\omega_b, \omega_b) \ket{\omega_b, \omega_b}, 
\end{eqnarray}
where the first(second) quantum number in the ket notation encodes the frequency of left(right) propagating photons, that is 
$ \ket{\omega_a,\omega_b} \equiv (\hat{a}_{\omega_a}^+ \hat{a}_{\omega_b}^{-})^\dagger \ket{0}$.

We can now characterize the entanglement of the two-qubit state defined by Eq.~\eqref{freqtwostates}. As entanglement measure we use the entanglement entropy defined as $S(\ket{\psi_{\rm PS}}) = -\Tr\left[\rho_1 \log_2 \rho_1 \right]$, where we defined the reduced density operator $\rho_1 = \Tr_{sys2}\left\{\ket{\psi_{\rm PS}}\bra{\psi_{\rm PS}} \right\}$, obtained by tracing out system 2 to the global state of Eq.~\eqref{freqtwostates}. In Fig.~\ref{Entanglement}(b) and (c) we analyze the behavior of the entanglement for different system parameters, considering the frequency distribution of Eq.~\eqref{Comegaspontaneous} with the Lorentzian coupling function of Eq.~\eqref{Lorentzg}. We take the filters to be centered around frequencies symmetric with respect to the atomic resonance, that is we fix $\omega_a = \omega_0 - \delta$ and $\omega_b = \omega_0 + \delta$ in Eq.~\eqref{freqtwostates}. In Fig.~\ref{Entanglement}(b) we show that the entanglement entropy grows with the  frequency-offset $\delta$, until it saturates to a plateau. This is consistent with the intuition gathered by looking at Fig.~\ref{spontaneous_FD}. Indeed,  the output frequency distribution is separable close to the resonance condition, while positive or negative correlations  can be observed far from resonance. To further characterize correlations, in Fig.~\ref{Entanglement}(c) we show the entanglement entropy $S(\ket{\psi_{\rm PS}})$ as a function of the ratio between the width $\beta$ of the coupling function and the total emission rate $\Gamma$. For $\beta/\Gamma=1$ we observe  a minimum of the entanglement entropy, as the frequency distribution is approximately separable [see Fig.~\ref{spontaneous_FD}]. For small values of $\beta/\Gamma$ [Fig.~\ref{Entanglement}(d)] the state is strongly correlated, and in the limit $\beta/\Gamma\to 0$ the state asymptotically approaches a Bell state 
$\lim_{\beta/\Gamma\to 0}\ket{\psi_{\rm PS}}= \ket{\psi^-}\equiv 
\frac{1}{2}\left(\ket{\omega_a,\omega_a} - \ket{\omega_b,\omega_b} \right)$. On the contrary, for large values of $\beta/\Gamma$ [see Fig.~\ref{Entanglement}(e)], the output photons are anticorrelated in frequency and the state asymptotically approaches the following Bell state $\lim_{\beta/\Gamma\to \infty}\ket{\psi_{\rm PS}} =  \ket{\phi^+}\equiv 
\frac{1}{2}\left(\ket{\omega_a,\omega_b} + \ket{\omega_b,\omega_a} \right)$. Therefore, quadratically-coupled atoms represent a compelling tool to generate frequency-entangled states, which are relevant for a variety of quantum-information applications~\cite{PhysRevA.105.052429,lyons2018attosecond, chen2019hong, Fabre21}. In the following section, we provide another striking example of potential applications.
\section{Ideal controlled-phase gate for propagating photons}
\label{Control-Phase_Gate}

Let us now provide a compelling application of this phenomenology in quantum computing tasks. We consider a controlled-phase gate implemented on qubits defined with a dual-rail encoding, which is of increasing relevance for quantum computing with superconducting circuits~\cite{teoh22}. In particular, we consider an encoding where each qubit is defined as a single photon distributed along two propagating modes $\ann{a}$ and $\ann{b}$, such that the logical states are encoded as $\ket{0}_L = \crea{a} \ket{0,0} =  \ket{1,0}$ and $\ket{1}_L = \crea{b} \ket{0,0} = \ket{0,1}$. Regardless of the physical encoding, a c-phase (or control-Z) gate is defined by the transformation
\begin{align}\label{encoding}
&\ket{0_c}\ket{0_s}\rightarrow \ket{0_c}\ket{0_s}\nonumber\\\quad&
    \ket{0_c}\ket{1_s}\rightarrow \ket{0_c}\ket{1_s}\nonumber\\\quad&
    \ket{1_c}\ket{0_s}\rightarrow \ket{0_c}\ket{1_s}\nonumber\\\quad&
    \ket{1_c}\ket{1_s}\rightarrow -\ket{1_c}\ket{1_s},
\end{align}
where $c$ and $s$ denote the control and signal logical qubits, respectively.

It has been shown in Ref.~\cite{nogo_theorem} that, by first principles, it is impossible to implement this gate with unit fidelity using a single (point-like) two-level quantum emitter as the nonlinear element. The reason is that a distortion of the wavefunction frequency distribution is unavoidably introduced, as conflicting requirements on the input state should be imposed for the single- and two-photon scattering events. This no-go theorem has been derived assuming standard linear interactions. There are alternative proposals to circumvent such an intrinsic limitation, but they come at the cost of using multi-level artificial atoms~\cite{Zheng13}, an array of many emitters~\cite{Brod16}, or an active time-dependent modulation of the coupling strength~\cite{Mikkel20}.

\begin{figure}
\center
\includegraphics[width=.99\linewidth]{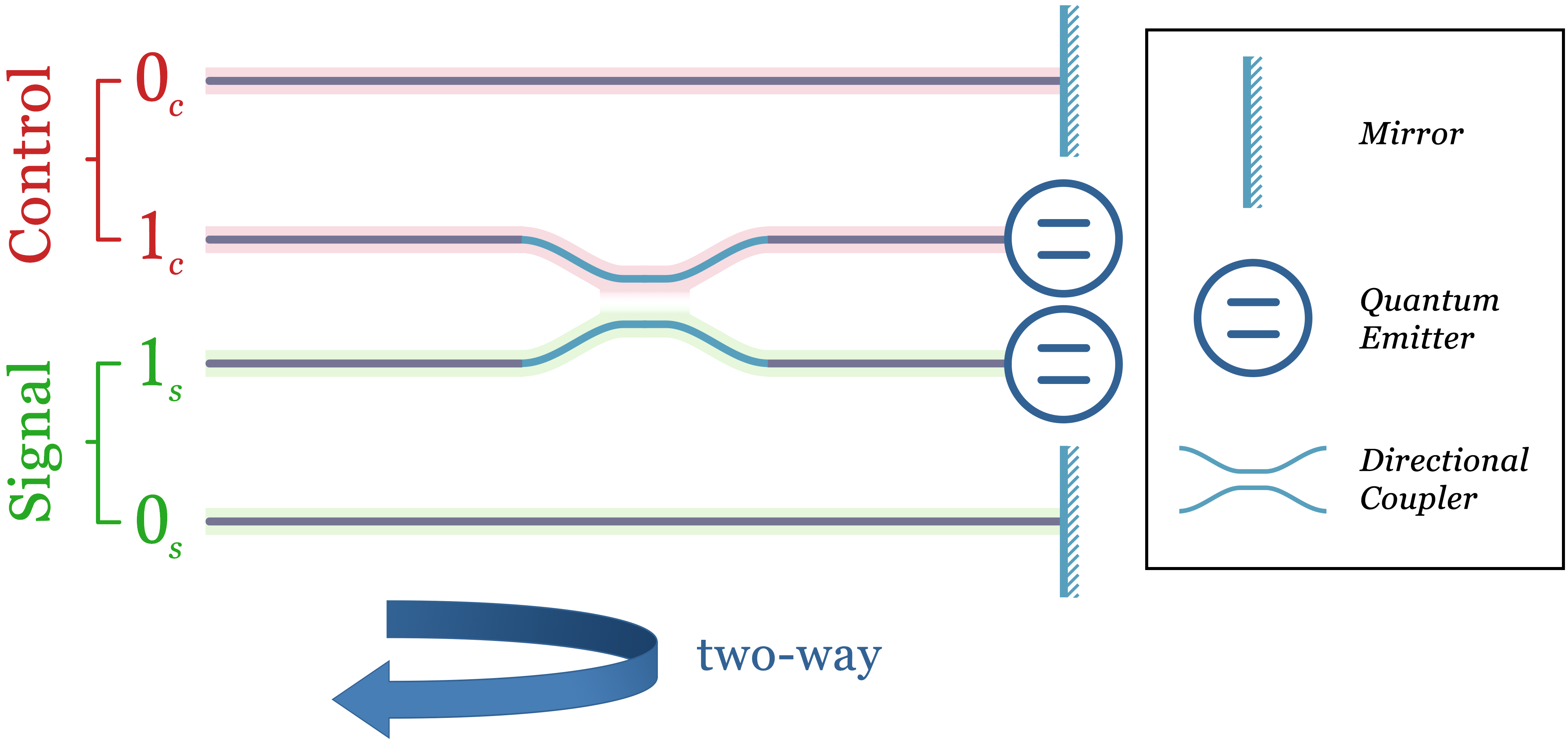}
    \caption{Sketch of the circuit scheme that implements a controlled-phase gate using two-level quantum emitters. The scheme has been proposed in ~\cite{nogo_theorem} to show intrinsic limitations to the achievable fidelity. Here, we show that in the case of quadratic interactions, there are no fundamental limits and the fidelity can in principle be arbitrarily close to 1. The origin of this advantage is that, for quadratic couplings, only the input state \(\ket{1_c,1_s}\) interacts with the emitters. For all other input states, only single photons impinge on the emitters, which are fully transparent in the one-photon subspace.}\label{graficogate}
  \end{figure}

In the following, we propose a straightforward scheme that can overcome this limitation and implement a passive protocol for a c-phase gate with (in principle) unit fidelity, by using a single quadratically coupled two-level emitter per path. The scheme (sketched in Fig.~\ref{graficogate}) is composed of four semi-open waveguides (two for the control and two for the signal photons), each encoding a state of one of the two logical qubits. A balanced beam-splitter (BS), represented as a directional coupler in the sketch, connects the $\ket{1_c}$ and $\ket{1_s}$ paths. As the BS is traversed back and forth, it does not modify the logical input states. As a result, since quadratically coupled emitters are transparent to single-photon signals, it can be shown by direct inspection that the first three operations of Eq.~\eqref{encoding} are correctly implemented. When the input state is $\ket{1_c,1_s}$, after the first passage in the BS the state will be $\frac{1}{\sqrt{2}}\left(\ket{2_c,0_s} + \ket{0_c,2_s}\right)$, i.e. the photons will come out from the same path. The role of the BS is then to send both photons toward the same emitter if and only if the input state is $\ket{1_c,1_s}$.

Let us analyze the process in which two photons $\ket{2_c,0_s}$ (or equivalently $\ket{0_c,2_s}$) scatter from the quadratically coupled emitters. We will compute which wavefunction of the two input photons maximizes the gate fidelity, studying the problem in frequency space. When the emitter is placed at the end of a semi-infinite waveguide, all photons are reflected, and the probability amplitudes of the output field are given by Prop.~\ref{prop1} with $\gamma^{--}=\gamma^{-+}=0$, and the only non-zero coupling term is $g^{++}_\Delta=\sqrt{\frac{\gamma^{++}}{2\pi}} u(\Delta)$, with $\|u\|_{l^2}/\sqrt{2}=1$. From Eq.~\eqref{Gammagrande}, it follows that $\Gamma = \gamma^{++}$. As shown in Sec.~\ref{Two-photon_Scattering}, the optimal interaction is achieved when the input state wavefunction in frequency space matches the distribution of the coupling. Consequently, we assume a biphoton input state given by $C^{++}_{\overbar{\omega}\Delta}(t_0)=f(\overbar{\omega})u(\Delta)^*$, where $\| f\|_{l^2}=1$. 
Following Prop~\ref{prop1}, the output field is
\begin{equation}\label{Coutnonchiral}
    C^{}_{\overbar{\omega}\Delta}(t_1)=e^{-i\omega_0(t_1-t_0)}C^{}_{\overbar{\omega}\Delta}(t_0)\left[1-\frac{\Gamma}{\Gamma/2+i(\omega_0-\overbar{\omega})}\right],
\end{equation}
where we dropped the propagation-direction indexes for the sake of brevity. Note how the output field equals the input field $C^{++}_{\overbar{\omega}\Delta}(t_0)=f(\overbar{\omega})u(\Delta)^*$ times a Lorentzian (in square brackets) in $\overbar{\omega}$, which in the limit of $\overbar{\omega}\simeq \omega_0$ approaches $-1$. We may use this limit to implement a c-phase gate, by imposing that the input state $f(\overbar{\omega})$ is a Bell-shaped function, with FWHM $\delta$, concentrated around the resonance frequency $\omega_0$. In the limit of narrow photons, $\Gamma\gg\delta$, the output field acquires the desired nonlinear phase $C^{}_{\overbar{\omega}\Delta}(t_1)\approx-e^{-i\omega_0(t_1-t_0)}C^{}_{\overbar{\omega}\Delta}(t_0)$, up to a phase $e^{-i\omega_0(t_1-t_0)}$ common to all logical input states. 

\begin{figure}
\center
\includegraphics[width=.87\linewidth]{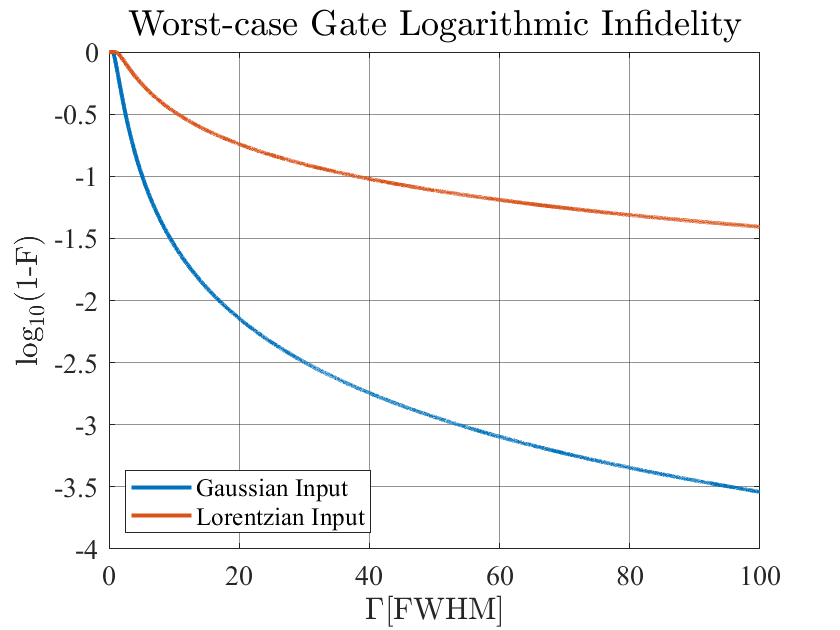}
    \caption{Worst-case gate logarithmic infidelity as a function of \(\Gamma\), expressed in units of the input FWHM, in the case of a chiral waveguide with coupling $g^{++}_\Delta=g^{++}u(\Delta)$. The input amplitude probability is chosen as $C^{++}_{\overbar{\omega}\Delta}=f(\overbar{\omega})u(\Delta)^*$. We compare two different relevant inputs: 
    $f(\overbar{\omega})$ real Normal distributed (blue) and $f(\overbar{\omega})$ real Lorentzian distributed (orange). Both the input frequency distributions \(f(\overbar{\omega})\) are centered in \(\omega_0\) and have the same FWHM. We see how in both cases the worst-case gate fidelity approaches unity. In particular, we observe a steeper slope for the Normal distributed \(f(\overbar{\omega})\) with respect to the Lorentzian one.}\label{graficofidelity}
  \end{figure}

Let us now analyze the performance of the controlled-phase gate, estimating the worst-case fidelity over all possible logical qubit states \(\ket{\psi}=a\ket{0_c0_s}+b\ket{0_c1_s}+c\ket{1_c0_s}+d\ket{1_c1_s}\). In a loss-less waveguide, the worst-case fidelity is defined as~\cite{nogo_theorem}
\begin{equation}\label{fidelity}
    F=\min_{\ket{\psi}} \left|\bra{\psi}\hat{U}^\dagger\hat{\mathcal{E}}\ket{\psi}\right|^2,
\end{equation}
where \(\hat{U}\) is the ideal c-phase gate and \(\hat{\mathcal{E}}\) are the actual operation experienced by the single and biphoton states. In our setup, only the two-photon state experiences a nontrivial scattering. We may thus write
\begin{align}\label{gates}
&\hat{U}\ket{\psi}=a\ket{0_c0_s}+b\ket{0_c1_s}+c\ket{1_c0_s}-d\ket{1_c1_s}\nonumber\\\quad& \hat{\mathcal{E}}\ket{\psi}=a\ket{0_c0_s}+b\ket{0_c1_s}+c\ket{1_c0_s}-d\ket{\Tilde{1}_c\Tilde{1}_s},
\end{align}
with \(\abs{a}^2+\abs{b}^2+\abs{c}^2+\abs{d}^2=1\) and \(\ket{\Tilde{1}_c\Tilde{1}_s}\) given by Eq.~\eqref{Coutnonchiral}. Inserting Eq.~\eqref{gates} into Eq.~\eqref{fidelity}, we obtain a formula that only depends on the overlap \(\bra{1_c1_s}\ket{\Tilde{1}_c\Tilde{1}_s}\)
\begin{align}\label{fidelity2}
    F&=\min_{\ket{\psi}} \left|\abs{a}^2+\abs{b}^2+\abs{c}^2-\abs{d}^2\bra{1_c1_s}\ket{\Tilde{1}_c\Tilde{1}_s}\right|^2\nonumber\\\quad&=\min_{d} \left|1-\abs{d}^2(1+\bra{1_c1_s}\ket{\Tilde{1}_c\Tilde{1}_s})\right|^2.
\end{align}

In Fig.~\ref{graficofidelity}, we show the worst-case gate logarithmic infidelity as a function of \(\Gamma\) expressed in units of the input FWHM. We compared two different relevant input cases: $f(\overbar{\omega})$ real Normal distributed and $f(\overbar{\omega})$ real Lorentzian distributed. Both distributions are centered on the resonance frequency \(\omega_0\) and have the same FWHM. We observe a steeper slope for the Normally distributed $f(\overbar{\omega})$ with respect to the Lorentzian one, due to the higher variance of the latter. Most importantly, in the limit of narrow inputs, FWHM$\ll\Gamma$, the worst-case gate fidelity approaches 1, thus overcoming the intrinsic limitation~\cite{nogo_theorem} found when considering linear interactions. This unanticipated result is of high conceptual value, as part of a prototypical analysis based of an effective model. An assessment of the  fidelity achievable in practice should be done considering a specific device and keeping into account the corresponding main sources of dissipation and decoherence.

\section{Discussion on physical implementations}\label{discussion}
Let us now provide comments on possible physical implementations of the considered model. There are two main approaches to implement quadratic interactions in the context of cavity QED. The first one consists in designing \emph{effective} implementations, or analog quantum simulations, where an external driving is used to selectively activate pump-induced interactions. This approach can be applied to a broad variety of quantum platforms and it was considered to design quadratic interactions first for atomic systems~\cite{felicetti_spectral_2015,PhysRevA.97.023624,PhysRevA.99.032303,Cong_selective,Schneeweiss_2018, Dareau_2018} and then in solid state devices~\cite{Steele_sidebands,Wang2016,Munoz2018}. The second approach consists in engineering \emph{genuine} quadratic couplings, such that the nonlinearity of the interaction is intrinsic and the exchange of excitations is not mediated by  external drivings. Genuine implementations have been designed for superconducting quantum circuits~\cite{felicetti_two-photon_2018,PhysRevA.98.053859,bertet2005dephasing} and nanomechanical resonators~\cite{Zhou2006}. In particular, the most promising platform to implement the proposed controlled-gate is circuit QED~\cite{Blais_review}, where decoherence and noise sources can be typically controlled within the 1\%. In this framework, it was shown that it is possible to engineer a genuine quadratic qubit-cavity coupling~\cite{felicetti_two-photon_2018} while fully inhibiting the linear coupling. It was also shown that a similar implementation can even be pushed into the ultrastrong coupling regime~\cite{PhysRevA.98.053859}. Such proposals are based on a flux qubit inductively coupled to a SQUID device, where the latter is operated in a weakly-nonlinear regime and is well approximated by a harmonic mode.

\begin{figure}
\center
\includegraphics[width=.99\linewidth]{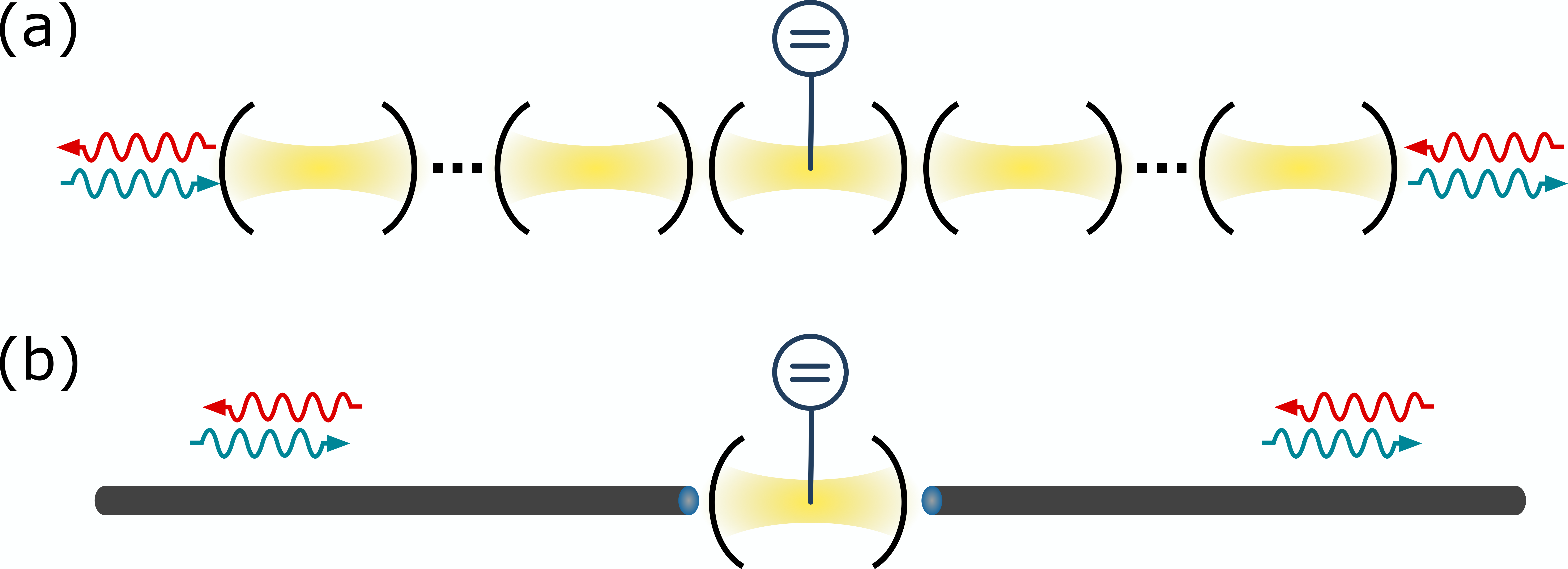}
    \caption{Sketch of proposed implementations of quadratic coupling with \emph{propagating} modes, generalizing established methods for the cavity QED setting. (a) A quadratically-coupled emitter embedded in an array of coupled resonators, that supports a discrete number of propagating modes. (b) A quadratic atom-cavity system linearly coupled to external waveguides. In the bad-cavity limit, the cavity mode can be adiabatically eliminated, obtaining an effective quadratic interaction between qubit and propagating modes.}\label{Implementations}
  \end{figure}

Currently, it is therefore well understood how to design genuine implementations of quadratic interactions between a quantum emitter and \emph{localized} harmonic modes. Going a step further, here we envision two ways in which these cavity QED settings could be directly generalized to implement quadratic couplings to \emph{propagating} modes: (i) The most straightforward generalization is shown in Fig.~\ref{Implementations}(a) and consists in implementing an array of coupled cavities which supports propagating modes~\cite{Ramos2016,Calaj02016}. A quadratically-coupled emitter could be added using known designs into one of the resonators of the array. In this case, the quadratic coupling function to propagating modes would depend on the collective mode structure of the resonator array, and in the weak coupling limit we expect the effective dynamics to be Markovian~\cite{Ramos2016} as assumed in this work. (ii) An alternative solution is shown in Fig.~\ref{Implementations}(b) and consists in considering a quadratic atom-resonator system in the bad cavity limit, where the resonator dissipation rate is large compared to the coupling so that its modes can be adiabatically eliminated~\cite{Auffeves2007,Reiter2012}. In this case, we expect an effective quadratic coupling with Lorentzian-like distribution between the artificial atom and the propagating modes of waveguides coupled to the leaky resonator.

These two solutions allow the implementation of the considered model with known results. A possible solution that would require further analyses consists in generalizing the superconducting circuit designs~\cite{felicetti_two-photon_2018,PhysRevA.98.053859} to a native waveguide-QED setting. In the simplest case, such a configuration could consist in a superconducting transmission-line waveguide that is galvanically coupled to a flux qubit through a SQUID. More complex configurations could be designed considering SQUID-based metamaterials. A detailed analysis of any of these implementations should be performed considering the microscopic modeling of the specific platform and this is outside the scope of this paper.

\section{Conclusions and outlook}
\label{Conclusions}
This work introduces the study of quadratic light-matter couplings in the field of waveguide QED. We have developed a general  scattering theory with a broad range of applicability to solid-state and atomic quantum technologies. The predicted phenomenology is fundamentally different from what is observable with standard dipolar couplings and it bears great potential for quantum information applications. Remarkably, we show that a single quadratically coupled emitter can overcome a well-known no-go theorem and implement a controlled-phase gate on flying qubits with unit fidelity. Let us discuss
the future perspectives in this novel research line, focusing first on the observation of an unconventional quantum phenomenology and then on the potential for quantum information applications.

In the framework of waveguide QED, quadratically-coupled emitters can be embedded in very different setups and configurations. For instance, the placement of artificial atoms in arrays with sub-wavelength spacing can be used to enhance/inhibit their individual emission rate and scattering properties, both in standard and chiral waveguides. The emergence of collective effects arising from many quadratically-coupled emitters has so far been studied only in the context of cavity QED~\cite{Delmonte_2ph_battery,Piccione2022}. Nonlinear waveguide QED devices can also undergo interaction-induced phase transitions~\cite{Iorsh20,Sedov20}. Alternatively, quadratically coupled emitters can be used as nonlinear mirrors that confine multiphoton states, by analogy with similar studies in the dipolar coupling regime~\cite{Chang_2012}. Furthermore, the quadratic coupling in waveguide-QED setups can be pushed into the non-Markovian or even in the ultrastrong-coupling regime, requiring the development of new theoretical and numerical tools~\cite{forn-diaz_ultrastrong_2017} and opening the door to new correlated phases and quantum phase transitions. Finally, nonlinear couplings can also be combined with the framework of giant atoms~\cite{kockum2021quantum}.

As demonstrated in this work, quadratic light-matter couplings are a compelling tool to process quantum information encoded in propagating photons. This possibility should be thoroughly addressed considering each specific experimental setting, with a detailed microscopic model of the system and realistic noise sources. The most interesting case is provided by superconducting quantum circuits, for which the dual-rail encoding is of high relevance for quantum computing~\cite{teoh22}.
 Finally, besides gates in the dual-rail encoding, quadratic light-matter couplings can be used to deterministically generate propagating biphoton states or to implement highly non-trivial photon subtraction processes. The possibility of tailoring the spectral features of emitted/scattered biphoton states paves also the way to the generation of entanglement in the time-frequency domain, which is relevant for quantum computation~\cite{PhysRevA.105.052429} and sensing~\cite{lyons2018attosecond, chen2019hong, Fabre21} applications.

\begin{acknowledgments}
We thank Francesco Ciccarello and Luca Leonforte for insightful discussions. U.A. and R.D. acknowledge support from the Academy of Finland, grants no. 353832 and 349199. Work in Madrid has been supported by the European Union's Horizon 2020 FET-Open project SuperQuLAN (899354), Proyecto Sinergico CAM 2020 Y2020/TCS-6545 (NanoQuCo-CM), and the CSIC interdisciplinary Thematic Platform (PTI+) on Quantum Technologies (PTI-QTEP+) T.R. further acknowledges support from the Ram\'on y Cajal program RYC2021-032473-I,
financed by MCIN/AEI/10.13039/501100011033 and the European Union NextGenerationEU/PRTR.
\end{acknowledgments}
 
\appendix

\begin{figure*}[ht!]
    \centering
    \includegraphics[width=0.90\textwidth]{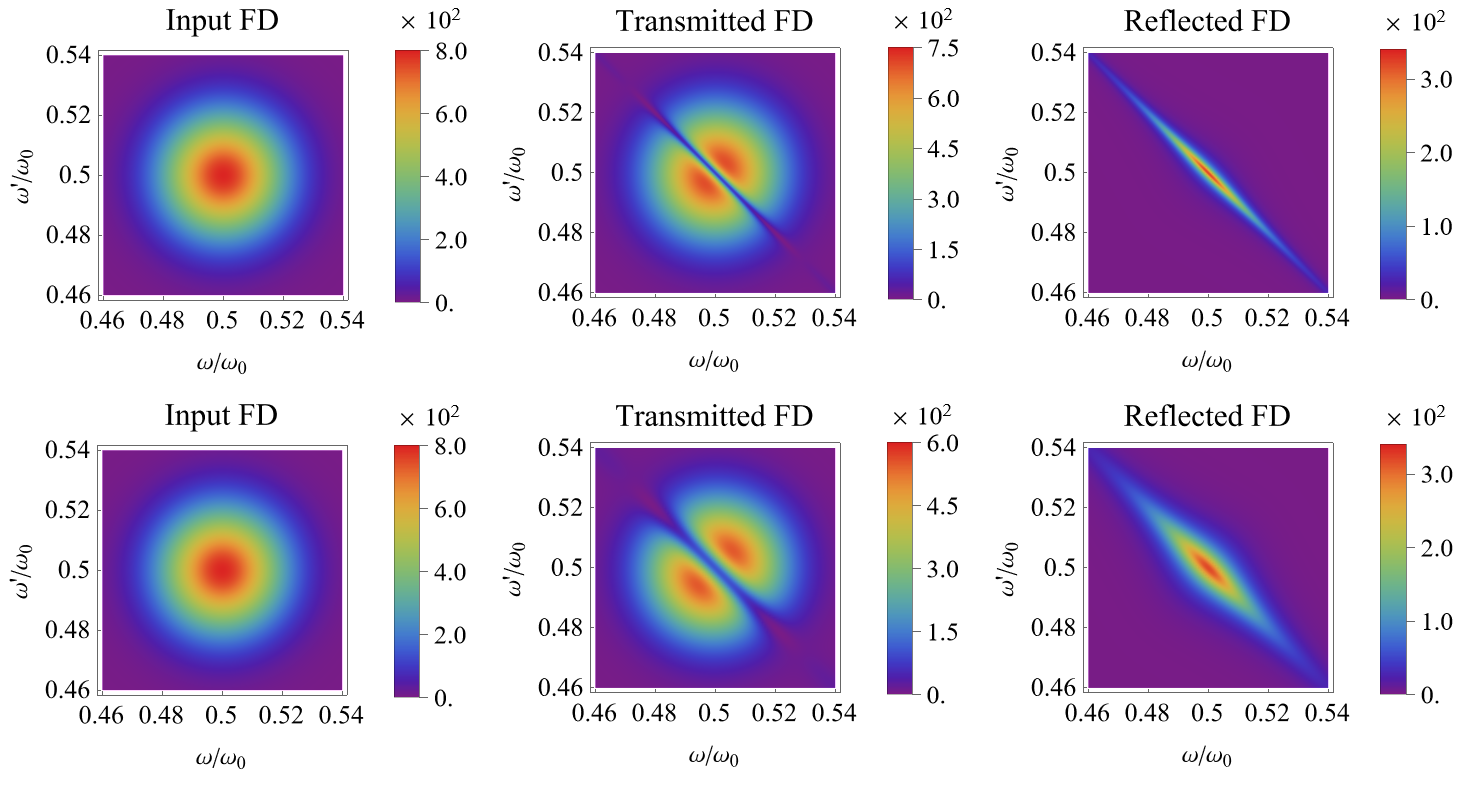}
    \caption{Two photon scattering, Lorentzian case.  We compare the transmitted and reflected/split FDs for two different values of the total emission rate, \(\Gamma = 0.004\omega_0\) for the first and \(\Gamma=0.012\omega_0\) for the second line. The FD for the reflection and splitting processes are identical. The FDs are shown in function of the two frequency variables $\omega$, $\omega'$. The input field  has been chosen as the product of two independent Gaussian FDs in \(\overbar{\omega}\) and \(\Delta\), centered at \(\omega_0\) (corresponding to the two-photon resonance condition), variance \(\alpha^2=(0.02\omega_0)^2\) and FWHM\(=2\alpha\sqrt{2\ln2}=0.047\omega_0\).  We consider a Lorentzian distributed coupling as in [Eq.~\eqref{Lorentzg}] with \(\beta=2\alpha\sqrt{2\ln2}\), in order to have the same FWHM for the input field and for the coupling.  We assume an isotropic spontaneous emission rate, i.e. \(\gamma^{\mu\mu^\prime}=\Gamma/4\). 
    }
    \label{outputFDLorentzian}
\end{figure*}

\section{Wigner-Weisskopf Theory}

In this section, we provide a detailed derivation of the scattering matrix (Prop.~\ref{prop1}) starting from the system of coupled differential equations for the amplitude probability coefficients in Eq.~\eqref{system1} and Eq.~\eqref{system2}.
\subsection{Markovian approximation}
\label{appendix_self}
We start by solving formally the differential equation for the field amplitude probability coefficient \(C^{\mu\mu'}_{\omega{\omega'}}(t)\):
\begin{align}\label{Comega2}
    &C^{\mu\mu'}_{\omega\omega'}(t)=C^{\mu\mu'}_{\omega\omega'}(t_0) e^{-i(\omega+\omega')(t-t_0)}\nonumber\\\quad& -i(g^{\mu\mu'}_{\omega\omega'})^\ast e^{-i(\omega+\omega')(t-t_0)}  \int_{t_0}^{t} C_e(\tau)e^{i(\omega+\omega')(\tau-t_0)} d\tau.
\end{align}
Inserting this result in Eq.~\eqref{system1} allows us to decouple the system of equations and find a linear self-consistent differential equation for \(C_e(t)\):
\begin{align}\label{Catom2}
&i\dot C_e(t)=\omega_0 C_e(t)+\nonumber\\\quad&\sum_{\mu,\mu'=\pm}\int{ g^{\mu\mu'}_{\omega{\omega'}} C^{\mu\mu'}_{\omega{\omega'}}(t_0)}e^{-i(\omega+\omega')(t-t_0)}d\omega d\omega'+ \nonumber\\\quad&-i\sum_{\mu,\mu'=\pm}\int d\omega d\omega' \int_{t_0}^{t} | g^{\mu\mu'}_{\omega{\omega'}}|^2C_e(\tau)e^{i(\omega+\omega')(\tau-t)} d\tau.
\end{align}
 The last term in Eq.~\eqref{Catom2} can be further simplified.
 First, we change the integration variables defining $\overbar{\omega}=\omega'+\omega$, $\Delta=\omega'-\omega$ and $d\omega d\omega'=\frac{d\overbar{\omega}d\Delta}{2}$, so that
 \begin{equation}
\int_{0}^{\infty} d\omega  d\omega^\prime
 \rightarrow
 \frac{1}{2}\left[
 \int_{-\infty}^{0}d\Delta \int_{-\Delta}^{\infty}d\overbar \omega +
\int_{0}^{\infty} d\Delta \int_{\Delta}^{\infty} d\overbar \omega
 \right].
 \end{equation}
 Then we rewrite the integral as 
 \begin{align}
 \label{before_Markov}
I&= \sum_{\mu,\mu'\in\{\pm\}} \int d\omega d\omega' \int_{t_0}^{t} | g^{\mu\mu'}_{\omega{\omega'}}|^2C_e(\tau)e^{i(\omega+\omega')(\tau-t)} d\tau=\nonumber \\
&=\sum_{\mu,\mu'\in\{\pm\}} \int_{0}^{\infty}d\Delta
\int_{\Delta}^{\infty}d\overbar \omega
|g^{\mu\mu'}_{\overbar \omega \Delta}|^2 \int_{t_0}^{t}C_e(\tau) 
  e^{i\overbar{\omega}(\tau-t)} d\tau = \nonumber \\ 
  &=  \int_0^{t-t_0}C_e(t-\tau)\ K(\tau) d\tau,
\end{align}
where we have used $g^{\mu\mu'}_{\overbar \omega \Delta} = g^{\mu\mu'}_{\overbar \omega -\Delta}$ and we have defined the kernel
\begin{equation}
K(t) = \sum_{\mu,\mu'\in\{\pm\}}\int_{0}^{\infty}d\Delta
\int_{\Delta}^{\infty}d\overbar \omega
|g^{\mu\mu'}_{\overbar \omega \Delta}|^2 
  e^{-i\overbar{\omega}t}.
\end{equation}
So far the treatment is exact. Now we perform the Markovian approximation on the basis of two assumptions. First, we assume the coupling strength to be weak with respect to the emitter frequency $\omega_0$. First-order perturbation theory allows us to rewrite the solution of Eq.~\eqref{Catom2} as $C_e(t)\approx e^{-i\omega_0 t} S_e(t)$, where $S_e(t)$ is a slowly-varying function of time. Second, we point out that, if the coupling strength has a weak dependence on $\overbar{\omega}$, the Kernel $K(t)$ is a rapidly-decaying function. Note that indeed, in the limit in which the coupling is flat in frequency, the Kernel is proportional to a Dirac delta in time. So, if the coupling strength is perturbative and sufficiently smooth with respect to $\overbar \omega$, we can approximate $S_e(t)$ as a constant in the short time interval in which $K(t)$ is non-vanishing. Accordingly, we can set  $S_e(t-\tau) \sim S_e(t)$ in the time integral in Eq.~\eqref{before_Markov} and recombine the terms so that
\begin{align}   
\label{K_int}
I = C_e(t) \int_0^{t-t_0} K(\tau) e^{i\omega_0 \tau} d\tau.
\end{align}
Note that for any specific form of the coupling strength, the integral in the last equation is now explicit and it can be evaluated by analytical or numerical means. However, to have a general analytical expression, we can take a further step and extend to infinity the upper limit of integration in Eq.~\eqref{K_int}. This corresponds to assuming that the Kernel correlation length is zero (which is standard in the Markov approximation, as it is exactly true when the coupling strength is natively constant with respect to $\overbar \omega$). Under this limit, performing the time-integration in Eq.~\eqref{before_Markov}, we obtain
\begin{align}
I &= C_e(t) \sum_{\mu,\mu'\in\{\pm\}}\int_{0}^{\infty}d\Delta
\int_{\Delta}^{\infty}d\overbar \omega\,
|g^{\mu\mu'}_{\overbar \omega \Delta}|^2 \times \nonumber\\
&\times\left[
\pi\delta(\overbar \omega - \omega_0) - i PV\left(\frac{1}{\overbar\omega - \omega_0} \right)
\right] = \\
 &= \left(\frac{\Gamma}{2} - i\delta_e \right) C_e(t),
\end{align}
where we have defined the total emission rate as
\begin{equation}
\Gamma = \sum_{\mu,\mu'\in\{\pm\}} \int_{0}^{\omega_0}
2\pi\, |g^{\mu\mu'}_{\omega_0 \Delta}|^2 \, d\Delta.
\end{equation}
We have also defined the Lamb shift
\begin{equation}
\delta_e = \sum_{\mu,\mu'\in\{\pm\}}\int_{0}^{\infty}d\Delta \lim_{\epsilon\rightarrow 0} \left[ \int_{\Delta}^{\omega_0 -\epsilon} d\overbar \omega + \int_{\omega_0 + \epsilon}^{\infty}d\overbar \omega \right]
\frac{|g^{\mu\mu'}_{\omega_0 \Delta}|^2 }{\overbar\omega - \omega_0},
\end{equation}
which is a small shift in the emitter resonance frequency. Here and in the main text, we absorb $\delta_e$ in the definition of $\omega_0$, as in most cases the experimental characterization of the emitter frequency already includes the Lamb shift. 
Now that all the terms in Eq.~\eqref{Catom2} have been simplified, we are able to write a  compact expression for the atom amplitude probability differential equation [Eq.~\eqref{Cdot} in the main text],
\begin{align}\label{Cdot2}
    &i\dot C_e(t) =\omega_0 C_e(t) -i \frac{\Gamma}{2} C_e(t) \nonumber\\\
    &+\sum_{\mu,\mu'=\pm}
     \int_0^\infty d\overbar\omega \int_{0}^{\overbar\omega}d\Delta
    { g^{\mu\mu'}_{\overbar \omega \Delta} C^{\mu\mu'}_{\overbar{\omega}\Delta}(t_0)}e^{-i\overbar{\omega}(t-t_0)},
    \end{align}
where for convenience we have used the relation
 \begin{equation}
\int_{0}^{\infty} d\omega  d\omega^\prime
 \rightarrow
 \frac{1}{2}
 \int_0^\infty d\overbar\omega \int_{-\overbar\omega}^{\overbar\omega}d\Delta ,
 \end{equation}
and $g^{\mu\mu'}_{\overbar \omega \Delta}= g^{\mu\mu'}_{\overbar \omega -\Delta}$,
$C^{\mu\mu'}_{\overbar{\omega}\Delta}(t) =C^{\mu\mu'}_{\overbar{\omega}-\Delta}(t)$.

\subsection{Scattering states}
\label{appendix_scattering}

As done in Eq.~\eqref{Comega2}, we can formally solve for $C^{\mu\mu'}_{\omega\omega'}(t)$ setting the initial conditions for a time $t_1\gg t_0$:
\begin{align}\label{Comega3}
    &C^{\mu\mu'}_{\omega\omega'}(t)=C^{\mu\mu'}_{\omega\omega'}(t_1) e^{-i(\omega+\omega')(t-t_1)}\nonumber\\\quad& +i(g^{\mu\mu'}_{\omega\omega'})^\ast e^{-i(\omega+\omega')(t-t_1)}  \int_{t}^{t_1} C_e(\tau)e^{i(\omega+\omega')(\tau-t_0)} d\tau.
\end{align}
Subtracting Eq.~\eqref{Comega2} from Eq.~\eqref{Comega3}, we find an input-output relation for the field before ($t=t_0$) and after ($t=t_1$)  the scattering event:
\begin{align}
&C^{\mu\mu'}_{\overbar \omega \Delta}(t_1) = e^{-i\overbar\omega(t_1 - t_0)} C^{\mu\mu'}_{\overbar \omega \Delta}(t_0)+ \nonumber \\
& - i \left(g_{\overbar \omega \Delta}^{\mu \mu'}\right)^* e^{-i\overbar\omega t_1} \int_{t_0}^{t_1} C_e(\tau) e^{i\overbar \omega \tau} = \\
&= e^{-i\overbar\omega(t_1 - t_0)} C^{\mu\mu'}_{\overbar \omega \Delta}(t_0) - 
i\sqrt{\gamma^{\mu \mu'}}e^{i\overbar \omega t_1} u(\Delta)^* \tilde{C}_e(\overbar \omega), \nonumber
\end{align}
where we switched to the frequency-sum and difference variables and we have used that $g_{\overbar \omega \Delta}^{\mu \mu'} = \sqrt{\frac{\gamma^{\mu \mu'}}{2\pi}} u(\Delta)$. We now take the Fourier transform of Eq.~\eqref{Cdot2} [or Eq.~\eqref{Cdot}] and integrate by parts the left side, assuming that $C_e(t_{0/1}) = 0$. We obtain
\begin{equation}
\tilde{C}_e(\overbar \omega) = -\sum_{\mu \mu'}\frac{i \sqrt{\gamma^{\mu \mu'}} e^{i\overbar\omega t_0 }}{\frac{\Gamma}{2} + i (\omega_0 - \overbar\omega)} \int_0^{\overbar \omega} u(\Delta) C_{\overbar \omega \Delta}^{\mu \mu'} d\Delta,
\end{equation}
where we have used that, for $t_1\gg t_0$, we can approximate $\int_{t_0}^{t_1} e^{i\tau(\omega -\overbar \omega)} = 2\pi \delta(\omega-\overbar \omega)$.
Finally, for the output state we find 
\begin{align}
&C^{\mu\mu'}_{\overbar \omega \Delta}(t_1)=  e^{-\overbar \omega(t_1 - t_0)} 
\left\{ C_{\overbar\omega \Delta}^{\mu \mu'}(t_0) + \right. \\ \nonumber
& \left. 
-\sqrt{\gamma^{\mu \mu'}} u(\Delta)^* \sum_{\alpha \alpha'} 
\frac{\sqrt{\gamma^{\alpha \alpha'}} }{\frac{\Gamma}{2} + i (\omega_0 - \overbar\omega)} \int_0^{\overbar \omega} u(\Delta) C_{\overbar \omega \Delta}^{\alpha \alpha'}(t_0) d\Delta
\right\}.
\end{align}

In the main text, we consider the narrow-bandwidth limit and extend to infinity the range of the integration over $\Delta$ in the previous expression. This corresponds to assuming that the input distribution is non-vanishing only in a region around a central frequency which is of the order of $\omega_0$ (for two-photon interactions the input photons are resonant when the sum of their frequencies matches the emitter frequency). We also remind that $u(\Delta)$ is bell-shaped and non-zero only for $\Delta\ll \omega_0$. 

\section{Lorentzian case}\label{appendix_Lorentzian}

In Sec.~\ref{Two-photon_Scattering}, we have presented results for the scattering of two-photon states in the Gaussian case, which can be analytically solved and allows one to understand the fundamental features of the two-photon scattering problem on a quadratically-coupled emitter. Let us now show that those features are not modified in the case of Lorentzian coupling functions [see Eq.~\eqref{Lorentzg}], which are more frequently encountered in realistic physical systems. This case can be solved semi-analytically using the results of Prop.~\ref{prop1}, by numerically integrating Eq.~\eqref{scatteringinputoutput}. The results are presented in Fig.~\ref{outputFDLorentzian} and good qualitative agreement is found with the Gaussian case presented in Fig.~\ref{outputFD}.

\bibliographystyle{apsrev4-1}
\bibliography{References.bib}

\end{document}